\documentclass[11pt,letterpaper]{article}

\usepackage[bookmarks,colorlinks,breaklinks]{hyperref}
\hypersetup{urlcolor=blue, colorlinks=true, citecolor=green!50!black, linkcolor=blue}
\usepackage[letterpaper, left=1in, right=1in, top=0.9in, bottom=0.9in]{geometry}
\usepackage[utf8]{inputenc}
\usepackage[american]{babel}
\usepackage[normalem]{ulem}
\usepackage{amsmath, amssymb, cases, amsthm}
\usepackage{thmtools}
\usepackage[capitalize,nosort]{cleveref}
\usepackage[shortlabels]{enumitem}
\usepackage{mdframed}
\usepackage{bbm}
\usepackage{bm}
\usepackage{microtype}
\usepackage{xcolor}
\usepackage{makecell}
\usepackage{mathtools}
\usepackage{float}

\declaretheorem[numberwithin=section,refname={Theorem,Theorems},Refname={Theorem,Theorems}]{theorem}

\declaretheorem[numberlike=theorem]{lemma}

\declaretheorem[numberlike=theorem]{definition}
\declaretheorem[numberlike=theorem]{claim}
\declaretheorem[numberlike=theorem,style=remark]{remark}
\declaretheorem[numberlike=theorem,refname={Fact,Facts},Refname={Fact,Facts},name={Fact}]{fact}

\declaretheorem[numberlike=theorem,refname={`Lemma',`Lemmas'},Refname={`Lemma',`Lemmas'},name={`Lemma'}]{lemmaq}

\theoremstyle{definition}

\def\final{0}  
\def\iflong{\iffalse}
\ifnum\final=0  
\newcommand{\yonggang}[1]{{\color{blue}[{\tiny Yonggang: \bf #1}]\marginpar{\color{blue}*}}}
\newcommand{\danupon}[1]{{\color{red}[{\tiny Danupon: \bf #1}]\marginpar{\color{red}*}}}
\newcommand{\sagnik}[1]{{\color{green!50!black}[{\tiny Sagnik: \bf #1}]\marginpar{\color{green!50!black}*}}}
\newcommand{\todo}[1]{{\color{red}[{\tiny TODO: \bf #1}]\marginpar{\color{red}*}}}
\else 
\newcommand{\yonggang}[1]{}
\newcommand{\danupon}[1]{}
\newcommand{\sagnik}[1]{}
\newcommand{\todo}[1]{}
\fi  

\usepackage[lined,ruled,linesnumbered,noend]{algorithm2e}
\SetKwInput{KwData}{Input}
\SetKwInput{KwResult}{Output}

\newcommand{\polylog}{\mathrm{polylog}}

\newcommand{\set}[2][ ]{\{#2 \ifthenelse{\equal{#1}{ }}{ }{~|~#1}\}}

\newcommand{\tOh}{\widetilde{O}}

\newcommand{\SNum}{\mathsf{Before}}

\newcommand{\conn}{\kappa}
\declaretheorem[numberlike=theorem]{assumption}

\usepackage{tcolorbox}
\usepackage{caption2}
\def\all{all}
\newtcolorbox{construction}[2][]
{
	colframe = gray!50,
	colback  = gray!10,
	coltitle = gray!10!black,
	left*=0mm, 
	before skip = 10pt,
	after skip = 10pt,
	title    = \textbf{\space\space #2},
	#1,
}

\renewcommand{\partial}{\mathsf{N}}

	\title{Finding a Small Vertex Cut on Distributed Networks}

\author{}
\author{Yonggang Jiang\thanks{MPI-INF, Germany, \texttt{yjiang@mpi-inf.mpg.de}} \and Sagnik Mukhopadhyay\thanks{University of Sheffield, UK, \texttt{s.mukhopadhyay@sheffield.ac.uk}} }

\date{}

\begin{document}
	
	\begin{titlepage}
		\maketitle \pagenumbering{roman}
		
		\begin{abstract}
We present an algorithm for distributed networks to efficiently find a small vertex cut in the CONGEST model. Given a positive integer $\kappa$, our algorithm can, with high probability, either find $\kappa$ vertices whose removal disconnects the network or return that such $\kappa$ vertices do not exist. Our algorithm takes $\kappa^3\cdot \tilde{O}(D+\sqrt{n})$ rounds, where $n$ is the number of vertices in the network and $D$ denotes the network's diameter. This implies $\tilde{O}(D+\sqrt{n})$ round complexity whenever $\kappa=\polylog(n)$.

Prior to our result, a bound of $\tilde{O}(D)$  is known only when $\kappa=1,2$ [Parter, Petruschka DISC'22]. For $\kappa\geq 3$, this bound can be obtained only by an $O(\log n)$-approximation algorithm [Censor-Hillel, Ghaffari, Kuhn PODC'14], and the only known exact algorithm takes  $O\left((\conn\Delta D)^{O(\conn)}\right)$ rounds, where $\Delta$ is the maximum degree [Parter DISC'19]. Our result answers an open problem by Nanongkai, Saranurak, and Yingchareonthawornchai [STOC'19]. 
\end{abstract}
		
		\setcounter{tocdepth}{2}
		\newpage
		\tableofcontents
		\newpage
	\end{titlepage}
	
	\newpage
	\pagenumbering{arabic}
	
	\ifx\all\undefined
        \end{document}
    \fi

\ifx\all\undefined
\begin{document}
	\fi
	
		\section{Introduction}\label{sec:introduction}

	

For any undirected non-complete\footnote{For complete graphs, the problem is trivial so we ignore this case.} 
	graph $G=(V, E)$, a set $S\subseteq V$ is called a {\em vertex cut} if $G\setminus S$ contains at least two connected components, where  $G\setminus S$ is obtained by removing vertices in $S$ from $G$. In the {\em vertex cut} or {\em vertex connectivity} problem, we are given a positive integer $\conn$ and want to either find a vertex cut of size at most $\conn$ or to answer that such vertex cut does not exist. 
	Vertex cut is a fundamental graph property and computing it is one of the most basic problems in graph algorithms. For example, it quantifies the vulnerability of a communication network in terms of the minimum number of vertices whose failures can disconnect the network. 
	In the {\em sequential} model, this problem has been extensively studied over many decades (e.g. \cite{Kleitman1969,Tarjan72,HopcroftT73,EvenT75,Even1975,Becker1982,Linial1988,KanevskyR91,CheriyanT91,Nagamochi1992,Henzinger2000,Gabow2006,Georgiadis10,HenzingerRW20,NanongkaiSY19,forster2020computing,LiNPSY21}). For $\conn=1$, a linear-time algorithm via depth-first search was long known due to Tarjan \cite{Tarjan72}. For $\conn=2$, the linear-time algorithm was due to Hopcroft and Tarjan \cite{HopcroftT73}. For $\conn=\polylog(n)$, an $\tilde O(m\conn^2)$-time algorithm was recently discovered by \cite{NanongkaiSY19,forster2020computing}. 
	For other values of $\conn$, a reduction to maxflow by \cite{LiNPSY21} together with the very recent fast maxflow algorithm of \cite{ChenKLPGS22} led to an almost-linear time algorithm.

	To conclude, the vertex cut/connectivity problem is almost solved in the sequential setting. 
	However, when it comes to {\em distributed networks computing their own vertex cut}, much less is known. This is the case even when it wants to find a few (say, $\conn=2$) vertices whose failures might destroy its communication. 
	A distributed algorithm for finding a small vertex cut is the focus of this paper.

	\paragraph{Distributed Vertex Cut.}	
	We study computing the vertex cut problem in the CONGEST model of distributed networks. 
	%
	%
	%
	In this model, an undirected graph $G=(V,E)$ is given as the communication network. Two important parameters are $n:=|V|$ and $D$, the diameter of $G$. Time is divided into discrete rounds. In each round, each vertex can send an $O(\log n)$ bits message to each of its neighbors. After each round, each vertex can locally perform arbitrary computation and decide what to send in the next round. Initially, each vertex is given a specified input indicating some local information of the network (e.g. neighbors and weights of its incident edges). For the vertex cut problem, the input of each vertex is simply the set of its neighbors and integer $\conn$. After several rounds, all vertices are expected to terminate and generate the desired output. For the case of the vertex cut problem, we expect at most $\conn$ vertices to identify themselves as being in a vertex cut if such a cut exists; otherwise, every vertex knows that such a cut does not exist.
	%
	%
	The goal is to minimize the number of rounds before all vertices terminate.


	The CONGEST model is a standard model for studying basic graph algorithms in the message-passing distributed networks, e.g. minimum spanning tree (MST), shortest paths, min-cut, and approximate maxflow~\cite{GarayKP98,KuttenP98,PelegR00,Elkin06,SarmaHKKNPPW12,GhaffariK13,Nanongkai2014,GhaffariKKLP15-maxflow,ForsterN18,Elkin20a,GhaffariL18,Dory2021}.
	These problems typically admit a trivial lower bound of $\Omega(D)$; thus, the focus is usually on the dependency on $n$. A large number of graph problems were shown to require $\tilde \Theta(D+\sqrt{n})$ rounds, and this bound has become a gold standard.\footnote{Throughout, $\tilde O$, $\tilde \Omega$, and $\tilde \Theta$ hide $\polylog(n)$.} 
	%
	%
	Examples of such problems include MST \cite{GarayKP98,KuttenP98,PelegR00,Elkin06,SarmaHKKNPPW12}, approximate shortest paths \cite{LenzenP13,Nanongkai2014,HenzingerKN21}, approximate 2-edge connected spanning subgraph (2-ECSS) \cite{Dory18, DoryG19}, 
	tree packing \cite{Censor-HillelGK14} and approximate maxflow \cite{GhaffariKKLP18}.
	%
	%
	For cut-related problems, a line of work (e.g. \cite{SarmaHKKNPPW12,GhaffariK13,Nanongkai2014,Daga2019,Ghaffari0T20,Dory2021}) led to an $\tilde O(D+\sqrt{n})$ bound for computing {\em edge cut} $\lambda$ that holds even for weighted graphs \cite{Dory2021,MukhopadhyayN20}. The bound matches 
	%
	%
	the lower bounds from \cite{GhaffariK13,SarmaHKKNPPW12} (the lower bounds hold when $\lambda$ is large enough).%
	\footnote{\cite{SarmaHKKNPPW12} proved a lower bound of $\tilde \Omega(\sqrt{n})$ for computing weighted mincut on some graphs of diameter $D=\Theta(\log n)$. For the unweighted case, it follows from \cite[Theorem 6.4]{GhaffariK13} that for any  $\epsilon>0$, there is a lower bound of $\tilde \Omega(n^{1/2-\epsilon})$ some graphs with diameter $D=\tilde O(n^{1/2-3\epsilon})$ and edge cut $n^{2\epsilon}$.} 
	Moreover, when the edge cut $\lambda$ is small, better algorithms exist: For $\lambda\in \{1,2\}$, the problem can be solved in $O(D)$ time \cite{PritchardT11}. For other values of $\lambda$, there is a  $O((\lambda D)^{O(\lambda)})$ rounds algorithm\cite{Parter19}. (The last bound is small under a typical assumption that $D\ll n$.)
	%
	
	In sharp contrast with the above, 
	our understanding of distributed vertex cut is much less complete. 
	To the best of our knowledge, existing algorithms consist of
	\begin{enumerate}[nosep]
		\item an $O(D+\sqrt{n}\log^*(n))$-round algorithm that works only when $\conn=1$ \cite{Thurimella97},
		
		\item an $O(D+\Delta/\log{n})$-round algorithm that works only when $\conn=1$ \cite{PritchardT11} ($\Delta$ denotes the maximum degree),
		
		
		\item an $O(\log n)$-approximation $\tilde O(\sqrt{n}+D)$-round algorithm \cite{CensorHillel2014}, and
		
		\item an $O\left((\conn\Delta D)^{O(\conn)}\right)$-round algorithm \cite{Parter19},
		
		\item an $\tOh(D)$-round algorithm that works only when $\kappa=1,2$ \cite{parter2022near}.
	\end{enumerate}
	Thus, even to find $\conn=3$ vertices that can disconnect the network, the available solutions are to either 
	settle with a much bigger approximate solution of size $\Theta(\log n)$  \cite{CensorHillel2014a}
	or
	find an exact solution in $O\left((\conn\Delta D)^{O(\conn)}\right)$ time \cite{Parter19} which can be prohibitively slow for typical networks with large-degree ``hubs'' 
	(e.g. the star networks). 
	%
	In other words, even for $\conn=3$ we are already very far from the typical $\tilde O(\sqrt{n}+D)$-time exact algorithms!

	\paragraph{Challenges.} A fundamental difficulty in solving the vertex cut problem is its tight connection to {\em maxflow computation}. For example, while edge cut algorithms that are faster than solving maxflow were known in the sequential model for many decades (e.g. \cite{NagamochiI92, Karger00, MukhopadhyayN20, GawrychowskiMW20, GawrychowskiMW21}, 
	it was only very recently that a vertex cut algorithm that is as fast as solving maxflow (and not faster) was found~\cite{LiNPSY21}. 
	%
	The situation is even worse in the distributed setting. For example, consider the case where we know two vertices $s$ and $t$ such that removing $\conn$ vertices in $G$ would disconnect $s$ from $t$ (this is a basic case that all the state-of-the-art sequential algorithms have to solve \cite{LiNPSY21,forster2020computing,NanongkaiSY19}). When $\conn=O(1)$, one can solve vertex cut in linear time in the sequential model using the Ford-Fulkerson algorithm. In contrast, in the distributed setting we cannot even solve this case in the typical $\tilde O(\sqrt{n}+D)$ rounds because it generalizes the {\em distributed reachability problem}, whose best-known round complexities are $\tilde O(D+\sqrt{n}D^{1/4})$ \cite{ghaffari2015brief} and $\tOh(\sqrt{n}+n^{1/3+o(1)}\cdot D^{2/3})$ \cite{liu2019parallel}. 
	More generally, 
	the distributed setting poses an additional challenge for computing vertex cut because 
	there is no non-trivial maxflow algorithm available.\footnote{The exception is the approximate maxflow algorithm of \cite{GhaffariKKLP15-maxflow}. However, approximate maxflow was not known to be useful for solving vertex cut.}
	Thus, to design distributed vertex cut algorithms, one needs to overcome fundamental questions of whether one could avoid maxflow computations or develop maxflow algorithms specialized for solving vertex cut. 
	Since efficient maxflow algorithms are not available in many models of computation (e.g. graph streaming and parallel computing), answering these questions may lead to efficient vertex cut algorithms in other models as well.


	\subsection{Our Result}
	
	We show that, in $\tilde O(D+\sqrt{n})$ rounds, a distributed network can find up to $O(\polylog(n))$ vertices that can disconnect itself. More generally, our result is the following. 
	
	\begin{theorem}[Informal. See~\Cref{thm:mainformal} for a formal version.]\label{thm:main}
		There is a randomized algorithm in the CONGEST model that, with input $\kappa<n^{1/4}$ and undirected graph $G$, takes $\conn^3\cdot \tilde{O}(D+\sqrt{n})$ rounds and determine whether $G$ is $\kappa$-vertex-connected or not; if not, output the minimum vertex cut.\footnote{When $\kappa\ge n^{1/4}$, our running time guarantee becomes at least $\Omega(kn)$, which is quite bad, so we do not consider this case here. Also notice that by running Ford–Fulkerson algorithm from a sampled node to any other nodes in parallel, one can easily get a $O(kn)$ algorithm.}
	\end{theorem}
	
	
	%
	%
	%
	Our bound can be thought of as generalizing the $\tilde O(D+\sqrt{n})$ bound of \cite{Thurimella97} that works only when $\conn=1$ to any $\conn=O(\log n)$; however, the techniques we use are very different.
	It is sublinear in $n$ as long as $\conn\ll n^{1/6}$. 
	Our result answers an open problem from \cite{NanongkaiSY19}.
	%
	%
	%

	\subsection{Techniques}\label{subsec:techniques}

	%
	We provide a detailed overview of this framework and our algorithm in the next section. 
	Here, we discuss some challenges and techniques to overcome them that might be of independent interest.
	Our algorithm follows the framework used by the algorithms of \cite{NanongkaiSY19,forster2020computing} for solving vertex cut in $\tilde O(m\kappa^2)$ time in the sequential model, where $m$ denotes the number of edges. 
	These algorithms consider two types of vertex cuts of size $\kappa$ (assuming that they exist): a vertex cut that leads to a small connected component $C$ is called {\em unbalanced} and otherwise it is called {\em balanced}.

	To find these cuts, we have to execute some {\em maxflow} algorithms which keep finding {\em augmenting paths}. For an intuition, 
	suppose that there are $\kappa$ internally vertex-disjoint $(s,t)$-paths between two vertices $s$ and $t$. An augmenting path is an $st$-path that, together with the
 existing paths, let us create $\kappa+1$ internally vertex-disjoint $(s,t)$-paths. 
	(See \cref{fig:augmenting} for an example and \Cref{sec:overview} for a more detailed definition.)
	Finding an augmenting path is useful because we can show that it exists if and only if there is no vertex cut of size $\kappa$ that disconnects $s$ and $t$.
	%
	We now consider finding two types of cuts. Note that below we use `Lemma' for lemmas that are used to provide intuition and are not actually proven.


	

	\paragraph{Finding Unbalanced Cuts: Local Flows and Resolving Congestions  (\Cref{lem:overviewSolvecongestion,lem:solvecongestion}).} 
	To find unbalanced cuts, \cite{NanongkaiSY19,forster2020computing} use {\em local flow} algorithms. 
	%
	Like many maxflow algorithms, a local flow algorithm keeps finding augmenting paths to increase the flow size; however, under some conditions, it can find augmenting paths {\em without reading the whole input graph}. For example, for vertex cut, \cite{NanongkaiSY19,forster2020computing} use local flow algorithms to solve a problem where, given a vertex $s$ in the above small connected component $C$, the algorithms can find the cut vertices in time roughly the size of the connected component $C$ defined above (more precisely, the {\em volume} of $C$), which can be much less than the size of the whole input graph. 
	%
	By not reading the whole graph, we can execute multiple local flow algorithms in near-linear time in total. This feature plays a key role in designing many efficient sequential algorithms, e.g. finding balanced cuts \cite{SpielmanT13,SaranurakW19}), edge cut \cite{KawarabayashiT15,HenzingerRW20}, and
	dynamically maintaining expanders \cite{Wulff-Nilsen17,NanongkaiS17,NanongkaiSW17,SaranurakW19,ChuzhoyGLNPS20}.
	%
	
	Applying the above idea in the CONGEST model, however, requires solving the {\em congestion issue}:
	many augmenting paths from different executions may go through the same edge. 
	%
	For example, the sequential vertex cut algorithms of \cite{NanongkaiSY19,forster2020computing} need to compute $\Omega(n)$ local flows at some point, and we cannot rule out the case where all these executions require augmenting paths that share the same edge, which would cause $\Omega(n)$ rounds to modify all $\Omega(n)$ flows along these augmenting paths.
	
	Congestion is a fundamental issue in the CONGEST model (thus the name). It is typically avoided by not executing too many algorithms in parallel. However, for vertex cut, we do not know how to avoid this.
	As far as we know, the same issue also arose in the {\em distributed expander decomposition computation}  \cite{ChangPZ19-expander,ChangS19-expander},
	where 
	the authors use {\em PageRank} algorithms instead of local flow algorithms (both algorithms can be used to compute the expander decomposition in the sequential model). Then, they exploit the property of PageRank to show that there is not much congestion, thus the congestion issue can be avoided. 
	
	In this paper, we solve the congestion issue differently. Essentially, we show that even when there are huge congestions, $\Omega(1)$ fraction of the executions can still proceed. To show this, we prove the following (see \Cref{lem:overviewSolvecongestion,lem:solvecongestion} for detail). 
	We have up to $\Omega(n)$ executions of the local flow algorithm of \cite{forster2020computing} running in parallel. 
	Consider two augmenting paths $p_1$ and $p_2$ from two executions with sources $s_1$ and $s_2$. If $p_1$ and $p_2$ meet at some vertex $t$, then  there is a path $p$ either from $s_1$ to $s_2$ or from $s_2$ to $s_1$ that uses only edges explored by the two executions so far such that {\em $p$ can be used as an augmenting path by one of the two executions}.
	\footnote{Here, we also exploit the fact that a source of one execution can be a sink for other executions.}
	In other words, if the augmenting paths from two executions meet at the same vertex, then one of them can augment to another one. 
	%

	This argument can be extended to show that if many augmenting paths meet at a vertex, then they can stop and only use what they have explored to finish the augmentation for half of them. 
	This property helps reduce congestion when finding augmenting paths from different vertices. 

	To conclude, the above property allows us to find a vertex cut of size $\kappa$ in $\tilde{O}(\kappa^3\alpha)$ where $\alpha:=|C|$, the number of vertices in one of the connected components in the cut (see \Cref{lem:smallAlphaAlgorithm} for detail).
	Finally, note that given the prevalence of local flow algorithms in designing efficient graph algorithms,  similar issues to the above may arise for other problems, and it is interesting to see if our technique can be applied elsewhere.

	
	\paragraph{Finding Balanced Cuts: Specialized Fast Reachability Algorithm (\Cref{lem:reachability}).} Before discussing this case, note that the above algorithm with round complexity $\tilde O(\kappa^3\alpha)$ already lends itself to a sublinear time algorithm for vertex cut with $\kappa=O(1)$---one can use this algorithm for small $\alpha$, and Ford-Fulkerson and reachability algorithm when $\alpha$ is large. In order to improve the round complexity to $\tilde O(D+\sqrt{n})$ even when $\kappa=O(1)$, there is another fundamental barrier: the need to solve the {\em distributed reachability} problem. 
	
	For concreteness, assume that removing $\kappa=O(1)$ vertices leaves us with two connected components $A$ and $B$ each of $\Omega(n)$ vertices. This case cannot be solved efficiently by the local flow algorithm since $\alpha=\Omega(n)$. In the sequential setting, this case can be easily solved by sampling two vertices $s$ and $t$ and computing a $(s,t)$-maxflow of size $\Theta(\kappa)$ in a graph. To do this, simply find augmenting paths for $\Theta(\kappa)$ rounds (i.e. the Ford-Fulkerson algorithm). 
	This takes $O(m\kappa)$ time and succeeds with constant probability (since $Pr[s\in A \text{ and } t\in B]=\Omega(1)$). 
	In the CONGEST model, however, even answering a simpler question of whether there is {\em one} augmenting path from $s$ to $t$ (i.e., solving the $(s,t)$-\textit{reachability}) requires larger than $\tilde O(D+\sqrt{n})$ rounds: The best distributed algorithms for reachability require  $\tilde O(D+\sqrt{n}D^{1/4})$ rounds~\cite{ghaffari2015brief}  and $\tOh(\sqrt{n}+n^{1/3+o(n)}\cdot D^{2/3})$ rounds~\cite{liu2019parallel} . 
	
	
	In this paper, we develop an algorithm specialized for our case: when we want to find an augmenting path, we are solving a reachability problem where most edges in the graph are {\em undirected}.
	A result implied by our technique when $\alpha=\Omega(n)$ is as follows. (See \Cref{lem:reachability} for the full statement.)
	
	\begin{lemmaq}
		There exists a randomized CONGEST algorithm that, given two vertices $s,t\in V$ and a set of $\ell$ internally vertex-disjoint $(s,t)$-paths $P$, either returns an augmenting path or declares that such path does not exist. The algorithm takes $\ell^2\cdot \tOh(D+\sqrt{n})$ rounds.
	\end{lemmaq}
	

So, to find $\kappa$ internally vertex-disjoint $(s,t)$-paths, we use the above algorithm $\kappa$ times, taking $\kappa^3\cdot \tOh(D+\sqrt{n})$ rounds in total. This partially explains the round complexity of our final algorithm. 
	
	The main technique for proving the above `lemma' is to modify the framework in the reachability algorithms \cite{nanongkai2014distributed,ghaffari2015brief,liu2019parallel}: As usual, we sample hubs and grow BFS trees from each hub and build a virtual graph on the hubs. Our novelty is to use a clustering technique  (\Cref{lem:clustering}) to create a small number of \textit{strongly connected components} (or clusters) and give them some ordering with the following guarantee: Any vertex in a cluster can reach any vertex in another cluster which is ordered lower than the former cluster. This clustering lets us reduce the number of vertices and edges in the virtual graph without affecting reachability as well as makes it possible to broadcast the whole virtual graph. See~\Cref{subsubsec:largealpha} for an overview of this algorithm.
	
\subsection{Open problems}

This paper presents a study on the computational complexity of the vertex connectivity problem for small $\kappa$ in the CONGEST model. There are several avenues for future research that may further improve upon the findings presented in this study.

\paragraph{Vertex connectivity in CONGEST model.} 

\begin{itemize}
    \item (Small $\kappa$) 
    for small values of $\kappa$, it would be interesting to investigate whether it is possible to surpass the $O(D+\sqrt{n})$ running time with an algorithm given that there is no $\Omega(D+\sqrt{n})$ lower bound for unweighted vertex connectivity. Although algorithms have been developed that run in $\tilde{O}(D)$ rounds for $\kappa=1,2$, the true complexity for larger $\kappa$ remains unknown.

    \item (Large $\kappa$) the current best algorithms for the general vertex connectivity problem in the CONGEST model do not have sub-linear time complexity when $\kappa$ is as large as $\Theta(n)$. It would be interesting to explore the development of sub-linear algorithms for cases where $\kappa$ is large.

    \item (Universally optimal) In recent years, there have been many papers seeking universally optimal algorithms, starting from the work by Haeupler, Wajc and Zuzic \cite{Haeupler2021}. Since our algorithm meet the $\tilde{O}(D+\sqrt{n})$ upper bound for $\kappa=\text{polylog}(n)$, it would be interesting to explore the development of an algorithm that is universally optimal. 
    
\end{itemize}

\paragraph{Parallel vertex connectivity.} By combining the current best sequential algorithm for small $\kappa$ with the current best parallel algorithm for reachability with depth $n^{1/2}$, it is possible to develop an almost linear work parallel algorithm with depth $n^{3/4}$. It would be interesting to investigate whether it is possible to further reduce the depth of the algorithm to the best reachability algorithm depth of $n^{1/2}$ or better. As this paper provides an example of surpassing the reachability running time for small $\kappa$ in the CONGEST model, it is reasonable to expect that similar improvements may be possible in the parallel model as well.

\paragraph{Other models of computation.} In addition to advancements in the CONGEST and parallel models of computation, we would like to see further advancements in cut-query and two-party communication models, both in classical and quantum settings, for the problem of vertex connectivity (and minimum vertex cut). Notably, the edge connectivity (and minimum edge cut) has nearly been resolved within the classical setting \cite{RubinsteinSW18, MukhopadhyayN20, LeeLSZ21} and considerable progress has been achieved within the quantum setting \cite{LeeSZ21, ApersEGLMN22}. However, no substantial progress is made for vertex connectivity.

\paragraph{Other graph cut problems.} Ultimately, significant advancements have yet to be made in addressing alternative variants of graph cut problems, including directed edge connectivity and minimum weighted vertex cut, in CONGEST and other distributed models of computation. Consequently, any headway achieved in these domains within any of the distributed models of computation would be of considerable interest.


	\section{Overview}\label{subsec:overview}\label{sec:overview}

In this section, we sketch the proof of our main result, i.e. \Cref{thm:main}. For notations, we use the following: for $S\subseteq V$, $\partial(S)=\{v\mid \exists (u,v)\in E,u\in S,v\not\in S\}$ denotes the neighbors of $S$ in graph $G=(V,E)$, and $\partial^+(S)=\partial(S)\cup S$.

The crux of our algorithm is the subroutines called IsolatingSmallCut and SingleSourceLocalCut, which give guarantees as in  \Cref{lem:smallAlphaAlgorithm,lem:largeAlphaAlgorithm} below. 
We sketch the proofs of 	
\Cref{lem:smallAlphaAlgorithm,lem:largeAlphaAlgorithm} in \Cref{subsubsec:smallalpha} and \Cref{subsubsec:largealpha} respectively. Then, in \Cref{sec:proof-mainthm-techoverview}, we show how to combine them together by following the framework of \cite{forster2020computing}.

We also denote the vertex cut of the graph $G$ by $(L, S, R)$, where $|L| \leq |R|$ are the two sides of the cut, and $S$ is the set of vertices whose removal disconnects $L$ from $R$.  \Cref{lem:smallAlphaAlgorithm} roughly  guarantees that if we have a set of vertices $A\subseteq V$ such that, for some $\conn$-cut $(L, S, R)$, exactly one vertex in $A$ is in $\partial^+(L)$ (i.e. $L$ and its neighbors), then we will be able to find such a cut or a similar cut in $\tOh(\conn^3|L|)$ rounds. 
So, to find a small vertex cut $(L, S, R)$ when $|L|$ is small (the ``unbalanced case'' mentioned earlier), this algorithm will be fast assuming that we can find such an $A$.  
For intuition, note the following related sequential algorithms. 
{\bf (i)} In \cite{LiNPSY21}, the same statement to ours is proved in the sequential setting with an algorithm that takes max-flow time (which is currently almost linear \cite{ChenKLPGS22}). This is done via the {\em isolating cut technique} \cite{LiP20}, thus the word ``Isolating'' in the name of our algorithm. 
Unfortunately, we cannot use the same technique since we do not have an efficient exact max-flow algorithm in the distributed setting.
{\bf (ii)} In \cite{forster2020computing}, a similar statement can be guaranteed in $O(m\kappa^2)$ time in the sequential setting. Compared to our requirement that $|A\cap \partial^+(L)|=1$, the statement of \cite{forster2020computing} requires a weaker condition that $|A\cap L|\geq 1$ 
($A$ that satisfies this condition can be easily found, e.g. $A=V$). 
%
As we will show in \Cref{subsubsec:smallalpha}, our algorithm follows the idea of \cite{forster2020computing}, but our stricter condition
gives us some leverage to avoid the congestion issue that we would face if we simply followed the ideas of \cite{forster2020computing} (discussed in the previous section). 



	\begin{lemma}[IsolatingSmallCut($G=(V,E),A\subseteq V,\conn,\alpha$); Proof in Section~\ref{sec:smallAlpha}]\label{lem:smallAlphaAlgorithm}
		There exists a CONGEST algorithm that given an undirected graph $G=(V,E)$, a set of vertices $A\subseteq V$, and $\conn,\alpha\in\mathbb{N}$,\footnote{Every vertex knows of their membership in $A$ and $\conn,\alpha$.} either outputs a valid $\conn$-cut\footnote{Every vertex knows of their membership in $S$.} $(L, S, R)$ with one side $L$ such that $|A\cap \partial^+(L)|=1$, or outputs $\bot$. The output satisfies
		
		\begin{itemize}[noitemsep]
			
			\item if there exists a vertex set $L\subseteq V$ such that $|\partial(L)|<\conn,$ $|A\cap \partial^+(L)|=1,$ and $|L|\le\alpha$, then the algorithm outputs $\bot$ with at most constant probability\footnote{When we say ''with constant probability'' in this paper, we mean a constant less than 1.}, and
            
			\item the algorithm runs in $\tilde{O}(\conn^3\alpha)$ rounds. 
			
		\end{itemize} 
	\end{lemma}


\Cref{lem:largeAlphaAlgorithm} roughly guarantees that if we know two vertices $s$ and $t$ that are on the opposite sides of a $\conn$-cut, i.e. for some $\conn$-cut $(L, S, R)$ we have $s\in L$ and $t\in R$, then we can find a $\conn$-cut efficiently; here, ``efficiently'' means the dilation of $\tOh(\conn^{2.5}\sqrt{n}+\conn^3 D)$ and congestion of $\tOh(\conn^{2.5}|L|/\sqrt{n})$.
%
We need the congestion to be $\tOh(\conn^{2.5}|L|/\sqrt{n})$ so that we can run $O(n/|L|)$ algorithms with different $s,t$ simultaneously, while still keeping the running time $\tOh(\conn^{2.5}\sqrt{n}+\conn^3 D)$. It is necessary to run $\Theta(n/|L|)$ algorithms since we need to sample $\Theta(n/|L|)$ vertices to guarantee at least one vertex is inside $L$. Each algorithm will take one sampled vertex as $s$. 

Note that a similar statement was achieved in the sequential setting in $O(m\conn)$ time, which is the time to compute a max-flow of size $\conn$ using Ford-Fulkerson algorithm. As discussed earlier, computing a max-flow will not allow us to beat the time to solve reachability. For this reason, we need some clustering ideas which we show in \Cref{subsubsec:largealpha}.
	
	\begin{lemma}[SingleSourceLocalCut($G=(V,E),s,t,\conn,\alpha$); Proof in Section~\ref{sec:largeAlpha}]\label{lem:largeAlphaAlgorithm}
		There exists a CONGEST algorithm that given an undirected graph $G=(V,E)$, two vertices $s,t\in V$ and $\conn,\alpha\in\mathbb{N}$, where $\conn\le\alpha$, either outputs a valid $\conn$-cut, or outputs $\bot$, such that

		\begin{itemize}[noitemsep]
			\item if there exists $L\subseteq V$ such that $|\partial(L)|<\conn,\{s,t\}\cap \partial^+(L)=\{s\},|L|\le\alpha$, then the algorithm outputs $\bot$ with constant probability,
			\item the algorithm has dilation $\tOh(\conn^{2.5}\sqrt{n}+\conn^3 D)$ and congestion $\tOh(\conn^{2.5}\alpha/\sqrt{n})$.
		\end{itemize} 
		
	\end{lemma}


\begin{remark}\label{rem:one-sided-algo}
    Throughout this paper, it is important for the reader to keep in mind that our algorithm is a Monte Carlo algorithm with one-sided error. Specifically, when the output is a cut, it must be a valid cut with a size less than $\kappa$. However, when the output is $\bot$, it is possible that the graph has a cut with a size less than $\kappa$, and the algorithm \textbf{cannot} distinguish whether the output is correct or not. Nevertheless, since the algorithm has one-sided error, as long as the error probability is bounded by a constant between 0 and 1, it can be reduced to as small as $\frac{1}{n^c}$ by repeating the algorithm $O(\log n)$ times.
\end{remark}
\subsection{IsolatingSmallCut (proof sketch of \Cref{lem:smallAlphaAlgorithm}; full proof in \Cref{sec:smallAlpha})}\label{subsubsec:smallalpha}

The starting point is to run the algorithm of~\cite{forster2020computing} for every vertex in $A$ simultaneously, i.e, run $\conn$ rounds of DFS to find \textit{augmenting path} on residual graphs, defined below.
For a path $p=(v_0,v_1,v_2,...,v_\ell)$,
we define $pre_p(v_i)=v_{i-1}$ for any $0<i\le \ell$ and $suc_p(v_{i})=v_{i+1}$ for any $0\le i<\ell$.
$v_1,v_2,...,v_{\ell-1}$ are called the \emph{internal vertices} of $p$. A set of paths are called \emph{internally vertex disjoint} if any two of them do not share the same internal vertex.
We define $V(p)$ as the vertex set consisting of all vertices in $p$. For a set of paths $P$, we define $V(P)=\cup_{p\in P}V(p)$.
	
	\begin{definition}[$(G,s,P)$-Augmenting Path]\label{def:overviewaugmentingpath}
		Let $G=(V,E)$ be an undirected graph, $s\in V$ and $P$ is a set of $k$ internally vertex disjoint paths starting from $s$. (We call $P$ a \textbf{flow-path set} of $s$.) A path $p_{aug}$ in $G$ is called {\em $(G,s,P)$-augmenting} if \begin{enumerate}[label=(\roman*),noitemsep]
			\item \textbf{Starting vertex:} $p_{aug}$ starts at $s$ and, \label{itm:start-vertex}
			\item \textbf{Forced retreat:} for any consecutive vertices $u_1,u_2$ in $p_{aug}$ where $u_2$ is not the end of $p_{aug}$ and any $p\in P$, if $u_2\in V(p)\setminus \{s\}$ and $u_1\not=suc_p(u_2)$, then $suc_{p_{aug}}(u_2)=pre_p(u_2)$.
   \label{itm:forward-edge}
		\end{enumerate}
	\end{definition}
	
\begin{figure}[]	
		\centering
		\includegraphics[scale=1]{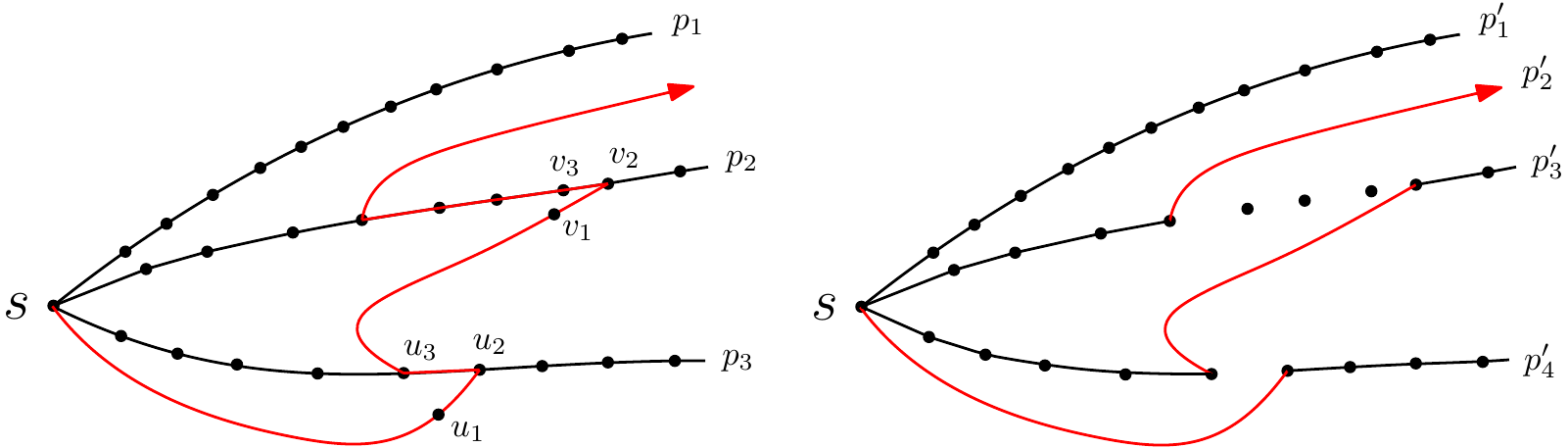}
		\setcaptionwidth{0.95\textwidth}
		\caption{\small The left figure is an example of \Cref{def:overviewaugmentingpath} with $P=\{p_1,p_2,p_3\}$. The red line starts at $s$. The consecutive three vertices $(u_1,u_2,u_3)$ satisfy $u_2\in V(p_3),u_1\not= suc_{p_3}(u_2),u_2\not=s$, so $u_3=pre_{p_3}(u_2)$. The same holds for $(v_1,v_2,v_3)$. Thus, the red line is a $(G,s,P)$-augmenting path. The right figure is the resulting $P'=\{p'_1,p'_2,p'_3,p'_4\}$ according to \Cref{lem:overviewaugmentingresult}.}\label{fig:augmenting}	
	\end{figure}

	\Cref{fig:augmenting} provides an example of such an $(G,s,P)$-augmenting path. Intuitively speaking, if an augmenting path enters a vertex in path $p\in P$ that is not from its successor, then it is forced to go backward (or \textit{retreat}).
	
	For a minimum vertex cut $(L,S,R)$, our goal is to find the maximum number of vertex disjoint paths from $s\in L$ to $R$ (from which we can infer the vertex cut), and we use augmenting paths to this end as follows. \Cref{lem:overviewaugmentingresult} shows (i) if an augmenting path ending at $R$ can be found, then we can increase the number of vertex disjoint path, (2) if no augmenting path ending at $R$ can be found, then we can find a vertex cut.
	
\begin{figure}[]	
		\centering
		\includegraphics[scale=1]{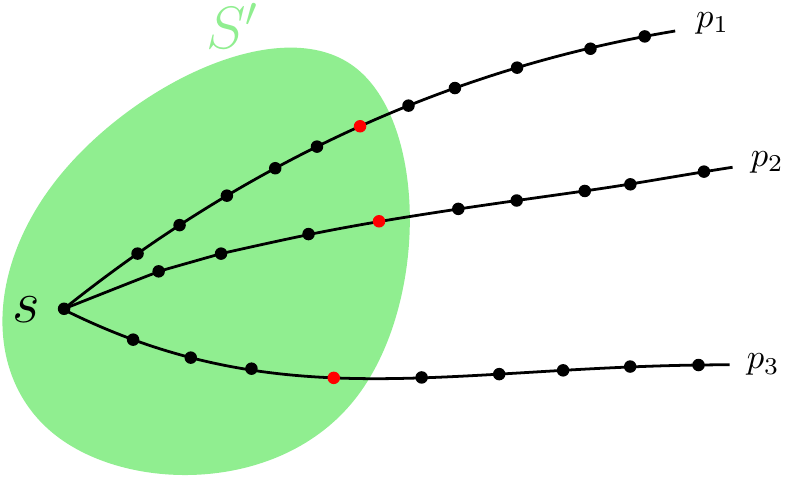}
		\setcaptionwidth{0.95\textwidth}
		\caption{\small $S'$ contains all the nodes that $s$ can reach through a $(G,s,P)$-augmenting path. The set of red vertices are vertices in $S'$ farthest from $s$ in each of their paths and form a vertex cut. }\label{fig:findcut}	
	\end{figure}

	
	\begin{lemmaq}[Simplified version of \Cref{lem:augmentingConnectivity,lem:correspondingCut}]\label{lem:overviewaugmentingresult}
		Suppose $G=(V,E)$ is an undirected graph, if $P$ is a set of $k$ internally vertex disjoint paths starting from $s\in V$, ending at a vertex set $T$, then
		\begin{enumerate}[label=(\roman*),noitemsep]
			\item (Augmentation.) Suppose $p$ is a $(G,s,P)$-augmenting path, ending at $t$, then there exists a set of $k+1$ internally vertex disjoint paths $P'$ ending at $T\cup\{t\}$. See~\Cref{fig:augmenting} as an example. 
			\item (Find a cut.) \label{item:lemfindcut}Let $S'$ contain all the nodes that $s$ can reach through a $(G,s,P)$-augmenting path. If $S'\neq V$, 
            then the following nodes form a vertex cut: for any $p\in P$, the node in $S'\cap V(p)$ that has the largest distance to $s$ on $p$ . See~\Cref{fig:findcut} as an example. 
		\end{enumerate}
	\end{lemmaq}
	
	It is not hard to see that, in the \textit{Augmentation} case above, the minimum vertex cut separating $s$ and $T\cup\{t\}$ has a size at least $k+1$ if $s$ does not have an edge to $T\cup\{t\}$---this follows from Menger's theorem. 
 
 Our algorithm IsolatingSmallCut for \Cref{lem:smallAlphaAlgorithm} works as follows. Initially, each node $s\in A$ has an empty flow-path set $P_s$. We run $\conn$ iterations where, in each iteration, we increase the size of $P_s$  by $1$ for each vertex $s\in A$: In each iteration, very informally, each vertex $s$ sends a DFS token to explore $G$ in a DFS manner for $\Theta(\conn\alpha)$ rounds in order to find a $(G,s,P_s)$-augmenting path. If the DFS gets \textit{stuck} (This is explained shortly.), then we use~\Cref{lem:overviewaugmentingresult} to find a cut. Indeed, our main challenge is to reduce congestion caused by all of these DFS traversals running in parallel. To this end, we exploit the following property of augmenting paths which is the main technical lemma of this subsection. We start with some definitions which provide the necessary context.

 A $(G,s,P)$-augmenting path $p_{aug}=(s,v_1,...,v_{\ell-1},v_{\ell})$ is called \emph{retreating} if there exists $p\in P$, such that $v_{\ell}\in V(p)\backslash\{s\},v_{\ell-1}\not=suc_p(v_{\ell})$, i.e., the only way to extend $p_{aug}$ to a $(G,s,P)$-augmenting path $(s,v_1,...,v_{\ell},v_{\ell+1})$ is to set $v_{\ell+1}=pre_p(v_{\ell})$. For example, in~\Cref{fig:augmenting}, the red $(G,s,P)$-augmenting path from $s$ to $u_2$ is retreating.
 A $(G,s,P)$-augmenting path is called \emph{non-retreating} if it is not \emph{retreating}.
	
	\begin{lemmaq}[Simplified version of  \Cref{lem:solvecongestion}]\label{lem:overviewSolvecongestion}
		For any undirected graph $G=(V,E)$, consider two vertices $u,v\in V$, and let $P_u$ and $P_v$ be flow-path sets of $u$ and $v$, respectively. Let $p_u$ and $p_v$ be non-retreating $(G,u,P_u)$- and $(G,v,P_v)$-augmenting paths, respectively. 
		If $p_u$ and $p_v$ end at the same vertex, then there exists a path $p$ on the subgraph of $G$ resulting from combining all edges of $P_u, P_v, p_u, p_v$ such that $p$ is either $(G,u,P_u)$-augmenting ending at $v$, or  $(G,v,P_v)$-augmenting ending at $u$.
	\end{lemmaq}

    \noindent
    With \Cref{lem:overviewSolvecongestion}, the algorithm becomes the following. Denote the \emph{flow-path set} of $s\in A$ as $P_s$. Initially $P_s=\emptyset$. Run the following procedure for $\conn$ iterations: In each iteration, we make sure that the size of $P_s$ increases by 1 for all $s \in A$.
	\begin{enumerate}[label=(\roman*), nosep]
	    	\item \textbf{Whole-graph DFS:} \label{itm:wg-dfs} In parallel, every vertex $s \in A$ sends a token (denoted by the $s$-token) to explore new vertices in $G$ in a DFS manner: Each vertex $u$ (including $s$), once receiving the token, finds out which of its neighbors is not explored yet by the $s$-token, and sends the $s$-token to one such unexplored neighbor. The DFS follows the forced retreat property described in \cref{def:overviewaugmentingpath}, i.e., when an $s$-token arrives at a vertex $u$ on a flow-path $p\in P_s$ not from $suc_p(u)$, then the token must be sent to $pre_p(u)$.\footnote{The astute reader may observe that this DFS traversal may visit a vertex on a flow-path path more than once because it is forced to do so by a \textit{forced retreat}. In Algorithm \ref{alg:augmenting}, however, we use a directed graph representation that will be defined in \Cref{def:augmentingPath} which avoids this problem.}
	    	The DFS traversal ends in either of the following three ways:

	    	\begin{itemize}
			\item If $s$ explores $\Theta(\conn\alpha)$ vertices or $s$ reaches another vertex $t\in A$, it stops.
			
			\item If two tokens from $u,v\in A$ meet at a vertex $t$, then they stop, form a pair $(u,v)$, and report this fact back to $u$ and $v$ through DFS trees. Denote the path from $u$ and $v$ to $t$ in the DFS trees by $p_u$ and $p_v$ respectively. Define subgraph $H_{(u,v)}$ as the subgraph formed by the union of edges in $p_u,p_v,P_u,P_v$. This graph will be used in the next step.
			
			If many tokens $u_1,u_2,\ldots,u_\ell$ meet at $t$, we pair them up $(u_1,u_2),\ldots$ to get subgraphs $H_{(u_1,u_2)}, \ldots$. In the case where $\ell$ is odd, $u_\ell$ is allowed to continue its DFS $t$ onward. 
			
			\item If $s$ finishes DFS (i.e., has explored all vertices it can reach) without exploring $\Theta(\conn\alpha)$ vertices and without reaching another vertex $t\in A$, output the small cut using~\Cref{lem:overviewaugmentingresult} \ref{item:lemfindcut}(If several vertices finish DFS, we just need to pick an arbitrary one.)
		\end{itemize}
		Let $(L,S,R)$ be the vertex cut as claimed in~\Cref{lem:smallAlphaAlgorithm}, i.e., $|S|<\kappa$, $A\cap(L\cup S)=\{s\}$ and $|L|\le \alpha$. Note that $s$ succeeds in finding a $(G, s, P_s)$-augmenting path that terminates in $R$ in the first case with a constant probability: (i) If $s$ explores $\Omega(\kappa\alpha)$ vertices, then a random vertex among the explored vertices is in $R$ with probability at least $1-\frac{1}{\Omega(\kappa)}$. So we can choose this random vertex as the terminating vertex of the augmenting path\footnote{A-priori we do not know if our chosen vertex is in $R$ or not. However, we show that, if the algorithm outputs a valid vertex cut in the end, it will be a cut of size at most $\conn$. See Remark \ref{rem:one-sided-algo}.}. (ii) If $s$ reaches $t\in A$ that $t\not=s$, then $t$ is the terminating vertex and $t\in R$. 
  
            Once the DFS traversals stop for every $s \in A$, we move to the next step.
	    	
	    	\item \textbf{Subgraphs DFS:} \label{itm:sg-dfs} For each pair $(u,v)$, $u$ and $v$ run DFS traversal on $H_{(u,v)}$. These DFS traversals in all $H_{(u,v)}$'s are run simultaneously using the random delay technique \cite{ghaffari2015near} to avoid congestion\footnote{According to \cite{ghaffari2015near}, running independent CONGEST algorithms simultaneously can be done using random delay in $\tOh(\text{dilation}+\text{congestion})$ rounds. See \Cref{lem:randomdelay} for more details.}. If $u$ find a $(G,u,P_u)$-augmenting path $p$ to $v$, it uses $p$ to increase the size of $P_u$ by $1$. Do the same for $v$. \Cref{lem:overviewSolvecongestion} guarantees that one of $u$ and $v$ will succeed in finding an augmenting path. 
	\end{enumerate}
	
    Note that executing Step \ref{itm:wg-dfs} and \ref{itm:sg-dfs} will increase $|P_s|$ for a constant fraction of $s \in A$ by \Cref{lem:overviewSolvecongestion}. We repeat these two steps $O(\log n)$ times to make sure $|P_s|$ increases for every $s \in A$.

	\medskip\noindent
	\textbf{Round complexity.} We first bound the round complexity for the two steps. One can see that Step \ref{itm:wg-dfs} runs in $O(\conn\alpha)$ rounds. The round complexity of Step \ref{itm:sg-dfs} depends on the dilation (i.e., the diameter of subgraph $H_{(u,v)}$) and congestion (i.e., the maximum number of $H_{(u,v)}$ for different pairs $(u,v)$ that shares the same edge) which we bound below. We crucially use the following fact: A $(G,s,P)$-augmenting path $p$ of length $\ell$ w.r.t. a flow-path set $P$ can increase the number of path edges in the new flow-path set by at most an additive factor of $\ell$. \footnote{\label{fact:Pi-subset}This observation follows directly from the following fact which is easy to see. Suppose $s\in A$ has flow-path set $P^i_s$ at the end of each iteration $i$ (We assume $P^0_s = \emptyset$), and consider the $(G,s,P^i_s)$-augmenting paths $p^1_s,p^2_s,...,p^{i}_s$ that are used to generate different $P^i_s$: Each $p^i_s$ is a $(G,s,P^{i-1}_s)$-augmenting path. We claim that the edges in $P^i_s$ is a subset of edges in $p^1_s,p^2_s,...,p^{i}_s$. Note that it might not be true that the set of edges in $P^{i-1}_s$ is a subset of the set of edges in $P^{i}_s.$}
            


	\begin{description}[noitemsep]
		\item[Dilation.] Note that each $p^i_s$, $i \in [\conn]$, is of size $O(\conn\alpha)$. From the fact stated above, it is straightforward to bound the size of $H_{(u,v)}$ (which is composed of $p_u,p_v,P_u,P_v$) by $\tOh(\conn^2 \alpha)$.

		\item [Congestion.] The number of $H_{(u,v)}$ that contain an edge $e$ is bounded by the number of times $e$ is visited by DFS traversals in Step \ref{itm:wg-dfs}, as $e$ can be included in some $H_{(u,v)}$ only after it is visited in any DFS traversal in Step \ref{itm:wg-dfs} by $u$ or $v$. Every edge $e$ is included in at most one DFS traversal in each round of Step \ref{itm:wg-dfs}. Since Step \ref{itm:wg-dfs} lasts for $\tOh(\conn \alpha)$ rounds in each of the $\conn$ iterations, an upper bound on the number of times $e$ is visited by DFS traversals in Step \ref{itm:wg-dfs} is $\tOh(\conn^2\alpha)$.
		
		
	\end{description}
	
	
	The total round complexity is $\conn \times O(\log n) \times \tOh(\conn^2\alpha) = \tilde O(\conn^3 \alpha)$: The first $\conn$ is the number of iterations, $O(\log n)$ is the number of times Step \ref{itm:wg-dfs} and \ref{itm:sg-dfs} are repeated in each iteration. See \Cref{sec:smallAlpha} for more details.



	\subsection{SingleSourceLocalCut (proof sketch of \Cref{lem:largeAlphaAlgorithm}; full proof in \Cref{sec:largeAlpha}.)}\label{subsubsec:largealpha}
	

		
	

	%
 
For intuition, note that a statement similar to \Cref{lem:largeAlphaAlgorithm} can be shown in the sequential setting \cite{NanongkaiSY19,forster2020computing} by running the Ford-Fulkerson algorithm. This algorithm runs for $\conn$ iterations where in each iteration it increases the amount of $st$-flow by one via an augmenting path. We follow this basic idea but need some modifications. First, in each of the $k$ iterations, we randomly select some {\em terminals}, where each vertex has probability $O(1/(\conn\alpha))$ to be the terminal. We allow the augmenting path to end at a terminal instead of at $t$. This suffices because if there exists a vertex cut $(L,S,R)$ such that $\{s,t\}\cap (L\cup S)=\{s\},|L\cup S|\le\alpha$ (thus $L$ satisfies the condition in the first bullet of \Cref{lem:largeAlphaAlgorithm}), a simple union bound shows that the random terminals on all $\conn$ rounds are in $R$ with constant probability. The algorithm for finding the augmenting path is stated as the following lemma. 
%
%
We will use this algorithm with $x=\kappa\alpha$.
%
%
Recall from \Cref{def:overviewaugmentingpath} the notion of {\em flow-paths} and $(G,s,P)$-augmenting path. 


	\begin{lemmaq}[RandomAugmenting$(G=(V,E),s,P,x)$]\label{lem:reachability}
		There exists a CONGEST algorithm called RandomAugmenting that takes an undirected graph $G=(V,E)$, two vertices $s,t\in V$, integer $x$ and a set $P$ of flow-paths of $s$ where each path in $P$ has length bounded by $O(x)$, as input and the algorithm either
            \begin{itemize}[label=-, nosep]
		    \item  outputs a vertex cut of size $|P|$, or
		    \item outputs a $(G,s,P)$-augmenting path with length bounded by $\tOh(x)$, either ending at $t$, or ending at a random vertex $\tilde{t}$, where $\Pr[\tilde{t}=v]=O(1/x)$ for any $v\in V$.
		\end{itemize}   The algorithm has dilation $\tOh(|P|^{1.5}\sqrt{n}+|P|^2D)$ and congestion $\tOh(|P|^{0.5}x/\sqrt{n})$. 
	\end{lemmaq}

	 To prove \Cref{lem:largeAlphaAlgorithm} using \Cref{lem:reachability}, our algorithm starts with  $P=\emptyset$. It proceeds in $\conn$ iterations, where in each iteration we find a $(G,s,P)$-augmenting path 
using \Cref{lem:reachability} with $x=\Theta(\conn\alpha)$ to increase the size of $P$ by $1$. Since $|P|<\conn$, one can see that the dilation is $\conn\cdot\tOh(|P|^{1.5}\sqrt{n}+|P|^2D)=\tOh(\conn^{2.5}\sqrt{n}+\conn^3 D)$ and the congestion is $\conn\cdot\tOh(|P|^{0.5}x/\sqrt{n})=\tOh(\conn^{2.5}\alpha/\sqrt{n})$, which is what we want in~\Cref{lem:largeAlphaAlgorithm}. The rest of this section is devoted to showing the proof idea of \Cref{lem:reachability}.

	\paragraph{Proof idea of \Cref{lem:reachability}.} 
 We first review the framework for distributed reachability algorithms used in \cite{nanongkai2014distributed,ghaffari2015brief,liu2019parallel}.  (We will modify this framework to find a $(G,s,P)$-augmenting path as guaranteed in \Cref{lem:reachability}.) 
%
This framework consists of
two phases, where the first phase is identical in all algorithms in \cite{nanongkai2014distributed,ghaffari2015brief,liu2019parallel}, and these algorithms differ in the second phase. Suppose we want to find a path from $s$ to $t$. The two phases are:
	
	\begin{enumerate}[label=(\roman*), wide]
	\item\textbf{Build a virtual graph.} Pick appropriate parameter $d$ (we will pick $d=|P|^{1.5}\sqrt{n}$ to prove~\Cref{lem:reachability}). Construct a virtual\footnote{By ``virtual'' it means that edges in the virtual graph might not be edges in the input network.} graph $G_{vir}=(V_{vir}, E_{vir})$ where $V_{vir}$ (also called set of \textit{hubs}) includes every vertex of $V$ with probability $1/d$ as well as $s$, and an edge $e=(h_1,h_2)$ is included in  $E_{vir}$ if the distance from $h_1$ to $h_2$ in $G$ is at most $d$. $E_{vir}$ can be constructed by constructing a BFS tree $T_{h}$ of depth $d$ from each vertex $h \in V_{vir}$ in $G$.
		\label{reachability:phaseI}
		
		\item\textbf{Reachability in the virtual graph.} Find all the hubs that $s$ can reach in $G_{vir}$, denoted by $H_r$ (the way to efficiently find $H_r$ differs by different algorithms). Now we claim that $\cup_{h\in H_r}T_h$ are all the vertices $s$ can reach in the original graph $G$. \label{reachability:phaseII}
	\end{enumerate}
	
	The correctness is guaranteed by the following arguments: since we sample hubs with probability $\frac{1}{d}$, the path from $s$ to a vertex $v$ contains hubs with distance $\tOh(d)$ one after another along the path, with high probability. Therefore, the hubs in the path form a directed path in the virtual graph, where the last hub in the path has distance $d$ to $v$ in $G$.

	\paragraph{Using reachability algorithm to find an augmenting path.} Our definition for augmenting path in~\Cref{def:augmentingPath} can be reformulated as a directed path in a directed graph, by the standard way of duplicating each vertex into in-vertex and out-vertex. See~\Cref{subsec:vertexresidualgraph} for more detail. Thus, we can use a directed graph reachability algorithm to find an augmenting path.
	
	However, directly applying this framework to prove~\Cref{lem:reachability} is not efficient as there can be $\Omega(n/d)$ BFS tree constructions that can lead to dilation $\Omega(d)$ and congestion $\Omega(n/d)$. Recall that in~\Cref{lem:reachability} we want dilation $\tOh(|P|^{1.5}\sqrt{n}+|P|^2D)$ and congestion $\tOh(|P|^{0.5}x/\sqrt{n})$, where $x$ can be much smaller than $n$. There is no way to set appropriate $d$ to satisfy both the dilation and congestion.
	To achieve a better dilation and congestion trade-off, we will only grow a BFS tree on fewer carefully chosen hubs instead of all $\Omega(n/d)$ hubs.


	\paragraph{Path centered clustering.} 
	The key idea to reduce the number of BFS tree constructions is a structure called \emph{path-centered clustering}. The details of this structure are described in \Cref{def:clustering}, and \Cref{lem:clustering} shows that we can efficiently construct this structure. 
	Here we give a simplified version of the structure. Note that the following definition is different from the definition in~\Cref{subsec:pathcenteredclustering}, because the following definition failed to satisfy~\Cref{cla:before} in some cases, which affect the correctness of our algorithm. However, it shows the general idea of the more complicated definition, so we use it for ease of explanation.
	
\begin{figure}[]	
		\centering
		\centering
		\includegraphics[scale=1]{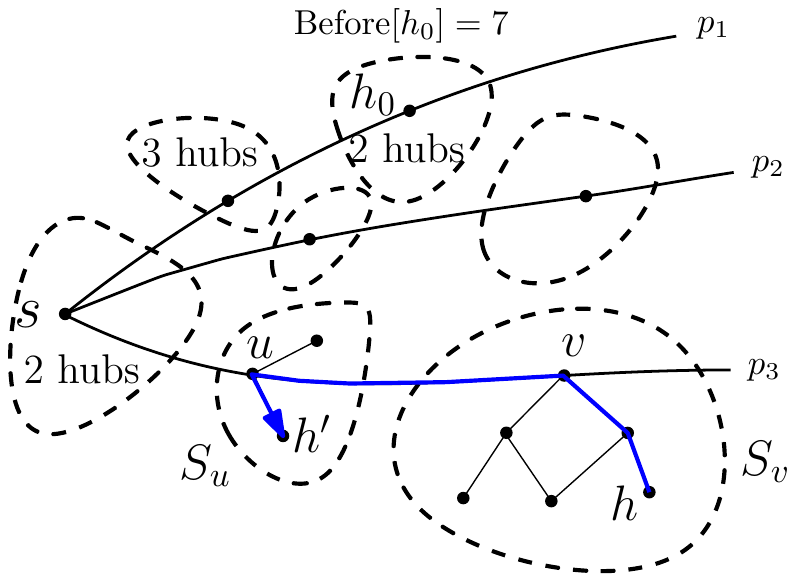}
		\setcaptionwidth{0.95\textwidth}
		\caption{\small Each dashed circle is a cluster. For simplicity, we only draw the inside structure of cluster $S_u$ and $S_v$. We can see that there are $7$ hubs $x$ with $x\preceq_{\mathcal{C}}h_0$, so $\SNum_\mathcal{C}[h_0]=7$. We also have $h'\preceq_{\mathcal{C}}h$. The blue line shows how $h$ can reach $h'$ through an augmenting path.}\label{fig:cluster}	
	\end{figure}

	For a given network $G=(V, E)$ of diameter $D$, a \emph{path centered clustering} is a tuple $\mathcal{C}=(P,\{S_u\}_{u\in V(P)})$ where $P$ is a flow-path set, and $\{S_u\}_{u\in V(P)}$ is a partition of $V$ (i.e. $V$ is a disjoint union of all $S_u$'s), called {\em clusters} with the following guarantees: Each cluster $S_u$ contains $u\in V(P)$, and each induced subgraph $G[S_u]$ has a diameter at most $D$. We call $u$ the {\em center} of every vertex $v\in S_u$ and denote it by $\mathsf{Center}_\mathcal{C}(v)$. See \Cref{fig:cluster} for an example. 
	
\paragraph{Definition of $\SNum$ and active hubs.}

We need a few definitions to show the properties of path centered clustering. For a path $p=(v_0,v_1,...,v_{\ell})$ and $v_i,v_j$ on the path, we say $v_i\preceq_pv_j$ if $i\le j$ and we say $v_i\prec v_j$ on path $p$ if $i<j$. For any two vertices $h,h'\in V$ and a path centered clustering $\mathcal{C}=(P,\{S_u\}_{u\in V(P)})$, we say $h'\preceq_\mathcal{C}h$, if $\mathsf{Center}_\mathcal{C}(h')$ and $\mathsf{Center}_\mathcal{C}(h)$ belong to some path $p \in P$ and $\mathsf{Center}_\mathcal{C}(h')\preceq_p \mathsf{Center}_\mathcal{C}(h)$. The relationship $\preceq_\mathcal{C}$ is not total as not every two vertices in $G$ are comparable by $\preceq_\mathcal{C}$. 
	For each hub $h$ (recall that hubs are sampled vertices in $G$ with sample probability $\frac{1}{d}$), we use $\SNum_{\mathcal{C}}[h]$ to denote the number of hubs $h'$ with $h'\preceq_{\mathcal{C}}h$. We will assume the following assumption.
	\begin{assumption}\label{cla:before}
	If $h'\preceq_{\mathcal{C}}h$, then $h$ can reach $h'$ through an augmenting path.
	\end{assumption}

	\begin{remark}\label{rem:canaugment}
		It is to be noted that our actual clustering is more fine-grained than what is described above to tackle the following technical problem: \Cref{cla:before} is true if $\mathsf{Center}_\mathcal{C}(h')\prec_p \mathsf{Center}_\mathcal{C}(h)$  (In~\Cref{fig:clustering}, the blue line shows an augmenting path from $h$ to $h'$.)
		and may not be true if $\mathsf{Center}_\mathcal{C}(h')= \mathsf{Center}_\mathcal{C}(h)$.
		This is solved by making the clustering more fine-grained---more details are provided in~\Cref{subsec:pathcenteredclustering}. In this section, we assume~\Cref{cla:before} holds for ease of explanation.
		
  
	\end{remark}
	
    \paragraph{Build a virtual graph with fewer BFS tree constructions.} In this part we will show how to build a virtual graph $G_{vir}$ on hubs with $O(x/d)\cdot |P|$ BFS tree constructions, such that either 
    
    \begin{itemize}[label=-, nosep]
        \item  $G_{vir}$ preserves the $s$-reachability (in the sense that all the vertices reachability by $s$ in $G$ can be reached from a vertex $u$ in $G_{vir}$ with distance $d$, such that $s$ can reach $u$ in $G_{vir}$), or 
	    \item $s$ can reach a random vertex $\tilde{t}$ such that each vertex in $V$ becomes $\tilde{t}$ with probability $O(\frac{1}{x})$.
	\end{itemize}
    
    Now we give our algorithm. We first compute a path centered clustering $\mathcal{C}$. We call a hub $h$ \emph{active hub} if $\SNum_\mathcal{C}[h]=O(x/d)$. Other hubs are called \emph{non-active hubs}. Denote the set of all active hubs as $V_{act}$. One can argue that $|V_{act}|=O(x/d)\cdot |P|$. We only grow BFS trees on active hubs. By setting $d=|P|^{1.5}\sqrt{n}$, the dilation and congestion of constructing all the BFS trees satisfy the requirement in~\Cref{lem:reachability}. By doing that, we can get a virtual graph $G_{vir}=(V_{vir},E_{vir})$ where $E_{vir}$ includes an edge $e=(h_1,h_2)$ if $h_1\in V_{act}$ and $h_1$ has distance at most $d$ to $h_2$ in $G$. 
   
    Now we argue the property of $G_{vir}$. If in $G_{vir}$, $s$ can reach a non-active hub $h$ through active hubs, then we can pick a uniform random hub $h'$ among all hubs $h'\preceq_{\mathcal{C}}h$ as the destination. Notice that a non-active node $h$ satisfies $\SNum_{\mathcal{C}}[h]=\Omega(n/d)$, thus, each node has probability at most $O(1/d)\cdot O(d/x)=O(1/x)$ to be the destination. On the other hand, if $s$ cannot reach any non-active hub, then by growing BFS trees on all active hubs, we can find all vertices that $s$ can reach in $G$ exactly.

	\paragraph{Find reachability in virtual graph} 
	Let $H_r$ contain all the active hubs that $s$ can reach in the virtual graph $G_{vir}$. Our goal in this part is to find $H_r$ efficiently. Notice that if we can find $H_r$, the according to the argument in the previous part, either we can find all vertices in $G$ that $s$ can reach, or find a non-active hub such that we can choose a random destination with probability $O(1/x)$.

	We first discuss the difficulty. Notice that $|H_r|=O(|P|\cdot x/d)=O\left(x/(|P|^{0.5}\sqrt{n})\right)$. Possible values of $x,|P|$ are $x=\Theta(n)$ and $|P|=O(1)$. In this case, $|H_r|=O(\sqrt{n})$. All the existing algorithms fail to find reachability with round complexity $\tilde{O}(\sqrt{n}+D)$ on a virtual graph with $\sqrt{n}$ vertices. However, our virtual graph is not an arbitrary directed graph. We will exploit some properties of our virtual graph to come up with an efficient algorithm.
	
	The idea is to sparsify the transitive closure of $G_{vir}$ and broadcast the whole sparsified graph. 
	We will make sure that the sparsified graph has the same reachability relationship as the original graph, and it is possible to broadcast the sparsified graph using $O(|P|\cdot |V_{act}|)$ messages. There are two types of edges in the sparsified graph. 
		\begin{description}[nosep]
			\item[Backward edges.] These are edges $(h,h')$ where $h'\preceq_{\mathcal{C}}h$. To learn this type of edge, we give each flow-path $p\in P$ an id. Each vertex $v$ on $p$ can learn $p$'s id and its position on $p$ (the number of vertices $x$ with $x\preceq_pv$) efficiently by existing results. After that, each active hub $h$ broadcasts the flow-path id where $\mathsf{Center}_{\mathcal{C}}[h]$ is on, as well as the position on the flow-path. 
			\item[Forward edges.] For each active hub $h$, recall that $T_h$ is the directed tree with depth $d$ rooted at $h$. Instead of keeping all edges from $h$ to all hubs in $T_h$, we preserve the ``highest hub'' for each path $p\in P$: let $T^p_h$ contain all hubs $x$ in $T_h$ with $\mathsf{Center}_{\mathcal{C}}[x]$ on $p$. Let $h^*_p$ be an arbitrary hub in $T^p_h$ such that for every other hub $h'\in T^p_h$, we have $h'\preceq_\mathcal{C}h^*_p$. $(h,h^*_p)$ is added to the virtual graph for any $p\in P$. 
			
		\end{description} 
		
		One can see that the number of messages broadcast by every active hub is bounded by $|P|$. Thus, the congestion is $\tOh(|P|^{0.5}x/\sqrt{n})$, which fits our goal. To see that the reachability relationship does not change, suppose $h'\in T_h$ where $\mathsf{Center}_{\mathcal{C}}[h']$ is on $p$, then $h$ can reach $h'$ in the virtual graph by first using the upward edge $(h,h^\ast_p)$, then using the downward edge $(h^\ast_p,h')$. 
		
		\begin{remark}
			We skip the mapping of each edge in the virtual graph to a path in the original graph efficiently in the technical overview, see \Cref{subsec:buildpartialvirtual graph} for more details. Actually, to recover the path in the original graph efficiently, the sparsified virtual graph defined in \Cref{subsec:buildpartialvirtual graph} is different from here and more complicated, while the high-level ideas are the same.
		\end{remark}

	\subsection{Putting everything together} \label{sec:proof-mainthm-techoverview}
	We first restate \Cref{thm:main} formally.
	
	\begin{theorem}\label{thm:mainformal}
		There is a randomized vertex cut algorithm in the CONGEST model that, with input $\kappa<n^{1/4}$ and undirected graph $G$, takes $\conn^3\cdot \tilde{O}(D+\sqrt{n})$ rounds, either outputs a minimum vertex cut of $G$, or outputs $\bot$, satisfying
		\begin{enumerate}
		    \item If the output is a vertex cut, then it must be a minimum vertex cut of $G$.
		    \item If $G$ is not $\kappa$-connected, then $\bot$ is output with at most constant probability. 
		\end{enumerate}
	\end{theorem}
	Since~\Cref{thm:mainformal} states a one-side error algorithm, the success probability can be boosted efficiently.
	The following is the schematic of the algorithm, using the subroutine described in \Cref{lem:smallAlphaAlgorithm,lem:largeAlphaAlgorithm}. 
	
	\begin{construction}{Schematic algorithm for vertex cut }
		\begin{itemize}
			\item \textbf{Input:} An undirected graph $G$ with $n$ nodes, a positive integer $\conn<n^{1/4}$.
			\item \textbf{Output:} A vertex cut with size less than $\conn$, or $\bot$. 
		\end{itemize}
		\begin{enumerate}
			\item If a vertex has degree less than $\kappa$ in $G$, output all the neighbors of this vertex. Otherwise continue the following procedures. 
			\item For $1\le i\le \log n$ do: 
			\begin{enumerate}
				\item 
				Let $\alpha=2^i,A=\emptyset$. Each vertex is included in $A$ with probability $1/\alpha$ independently.
				\item If $\alpha\le \kappa$, 
    \begin{itemize}
        \item discard vertices in $A$ with degree larger than $\kappa$ in $G[A]$, and run a $O(\kappa)$-coloring algorithm in $G[A]$ (\cite{HalldorssonKMT21}) to get $\ell=O(\kappa)$ independent sets $A_1,A_2,...,A_{\ell}$ (see~\Cref{cla:independentset});
        \item run IsolatingSmallCut$(G,A_i,\conn,\alpha)$ (see~\Cref{lem:smallAlphaAlgorithm}) for any $i\in[\ell]$.
    \end{itemize} 
				\item If $\kappa<\alpha<\sqrt{n}$, discard all vertices in $A$ with degree at least $1$ in $G[A]$, run IsolatingSmallCut$(G,A,\conn,\alpha)$ (see~\Cref{lem:smallAlphaAlgorithm}).
				\item If $\sqrt{n}\le \alpha$, for each $s\in A$, let $t_s\in A$ be an arbitrary vertex which is distinct from $s$. Run SingleSourceLocalCut$(G,s,t_s,\conn,\alpha)$ (see  Lemma~\ref{lem:largeAlphaAlgorithm}) for any $s\in A$ in parallel (see~\Cref{lem:randomdelay}). 
			\end{enumerate}
			\item If any subroutine described in \Cref{lem:smallAlphaAlgorithm,lem:largeAlphaAlgorithm} outputs a cut, then the algorithm outputs the cut and stop. Otherwise, output $\bot$.
		\end{enumerate}
	\end{construction}
	
	\paragraph{Correctness.} According to \Cref{lem:smallAlphaAlgorithm,lem:largeAlphaAlgorithm}, if a cut is output, then it must be a valid vertex cut with size less than $\kappa$. Thus, if the graph $G$ has no valid vertex cut with size less than $\kappa$, then the algorithm will output $\bot$ with probability $1$. 
	
	Suppose there is a vertex cut $(L,S,R)$ with $|S|<\conn$. We assume the max degree of the graph is at least $\conn$, otherwise, a vertex cut of size less than $\conn$ can be trivially found in the first step of the algorithm. We will show that in the second step, at the first iteration when $|L|<\alpha=O(|L|)$, a cut with a size less than $\conn$ will be output with constant probability. 
	\begin{description}[noitemsep]
		\item[Case 1 ($\conn \geq \alpha$):]In this case, we get $\ell$ independent sets $A_1,A_2,...,A_\ell$. We first prove the following lemma.
		
		\begin{lemma}\label{cla:independentset}
		    At least one of $A_1,A_2,...,A_\ell$ (denoted by $A^*$) satisfies: $A^*$ is an independent set on $G$, contains exactly one vertex in $L$, and $A^*\cap S=\emptyset$. 
		\end{lemma}
		\begin{proof}
			Since $|L|=\Theta(\alpha)$ and we sample each vertex into $A$ with probability $1/\alpha$, with constant probability there is exactly one vertex $u\in A\cap L$. Let $\partial^+(u)$ contain all neighbors of $u$ in $G$ and $u$ itself. Since the degree of $u$ is at least $\kappa$ and $|S|<\kappa$, we have $(L\cup S)-\partial^+(u)$ has size at most $|L|+\kappa-\kappa=|L|=O(\alpha)$. Thus, with constant probability, $(L\cup S)-\partial^+(u)$ contains no vertex in $A$. Consider the independent set $A^*$ among $A_1,...,A_\ell$ that contain $u$. We have $A^*\cap (L\cup S)=\{u\}$, which finishes the proof. 
		\end{proof}
		
		According to Lemma~\ref{lem:smallAlphaAlgorithm}, once~\Cref{cla:independentset} is proved, a cut with size less than $\alpha$ will be output with constant probability when IsolatingSmallCut$(G,A^*,\kappa,\alpha)$ is called.
		
		\item[Case 2 ($\kappa<\alpha<\sqrt{n}$):] Since we sample each vertex into $A$ with probability $1/\alpha$ and $|L|=\Theta(\alpha),S=O(\kappa)=O(\alpha)$, with constant probability, exactly one vertex is in $A\cap L$ and $A\cap S=\emptyset$. According to Lemma~\ref{lem:smallAlphaAlgorithm}, a cut with size less than $\alpha$ will be output with constant probability.
		\item[Case 3 ($\alpha \geq \sqrt{n}$):] According to the same argument, with constant probability, exactly one vertex $u$ is in $A\cap L$ and $A\cap S=\emptyset$. Consider the instance with $s\leftarrow u$, that instance satisfies the premise of Lemma~\ref{lem:largeAlphaAlgorithm} to output a cut with size less than $\conn$. 
	\end{description}
	
	\paragraph{Round complexity.} When $\alpha<\conn$, the round complexity for the coloring algorithm is $\tOh(1)$. There are $\tOh(\conn)$ instances of IsolatingSmallCut in Lemma~\ref{lem:smallAlphaAlgorithm}, which leads to the round complexity $\tOh(\conn^4\alpha)=\tOh(\conn^5)=\tOh(\conn^3\sqrt{n})$ since $\conn=O(n^{1/4})$. When $\conn\le\alpha<\sqrt{n}$, the round complexity is $\tOh(\conn^3\alpha)=\tOh(\conn^3\sqrt{n})$. When $\sqrt{n}\le\alpha$, the dilation is $\tOh(\conn^{2.5}\sqrt{n}+\conn^3D)$ and the total congestion is $\tOh(n/\alpha)\cdot\tOh(\conn^{2.5}\alpha/\sqrt{n})$, since there are $\tOh(n/\alpha)$ vertices in $A$ w.h.p. Thus, the round complexity is $\conn^3\cdot \tOh(\sqrt{n}+D)$.

\subsection{Organization} The rest of this paper is organized as follows. In \Cref{sec:preliminary}, we give some basic definitions and define the vertex residual graph. In \Cref{sec:smallAlpha}, we describe the algorithm IsolatingSmallCut to prove \Cref{lem:smallAlphaAlgorithm}. In \Cref{sec:largeAlpha}, we describe the algorithm SingleSourceLocalCut to prove \Cref{lem:largeAlphaAlgorithm}. Other proofs for less important lemmas are deferred to the appendices.

	\section{Preliminary}\label{sec:preliminary}
	\subsection{Basic Definitions}\label{subsec:basicdefinitions}
	We will use the following terminology throughout the paper.
	
	\paragraph{Graph terminologies.}For convenience, we treat an undirected graph as a directed graph with each undirected edge $(u,v)$ replaced by two directed edges $(u,v),(v,u)$, i.e., $(u,v)$ and $(v,u)$ are different edges in an undirected graph. 
	
	For a graph $G=(V,E)$, a \emph{path} $p$ with length $k$ is a vertex sequence $(v_0,v_1,...,v_k)$, where $(v_i,v_{i+1})\in E$ for all $0\le i<k$. We say $p$ starts at $v_0$ and ends at $v_k$. For $i<j$, we write $v_i\prec_pv_j$ to denote $v_i$ precedes $v_j$ on path $p$. The edges $\{(v_i,v_{i+1})\mid 0\le i<k\}$ are called the edges of $p$, denoted as $E(p)$. Normally we assume a path cannot contain repeated edges (but can contain repeated vertices). $\{v_i\mid 0\le i\le k\}$ are called vertices of $p$, denoted as $V(p)$, $\{v_i\mid 0< i< k\}$ are called internal vertices of $p$, denoted as $V_I(p)$. $p$ is called \emph{simple path} if $v_0,...,v_k$ are distinct. A set of paths $P$ are called \emph{internally vertex disjoint}, if every two paths intersect only at non-internal vertices (start and end vertices). Similarly we define $E(P)=\cup_{p\in P}E(p),V(P)=\cup_{p\in P}V(p),V_I(P)=\cup_{p\in P}V_I(p)$. We say $P$ ends at the multiset $V'$ if the union of end vertices of paths in $P$ is $V'$. In multiset, we also care about the number of occurrences of elements. A \emph{circle} with length $k$ is a length $k$ path $(v_0,v_1,...,v_k=v_0)$ where $v_0,v_1,...,v_{k-1}$ are different. A set of circles are called vertex disjoint if any two of the circles do not share any vertices.  
	
	A subgraph of $G$ is an edge set $E'\subseteq E$. Let $V'$ be all the vertices adjacent to $E'$, $(V',E')$ is the subgraph associated with $E'$. We do not distinguish $E'$ and the subgraph $(V',E')$ if there is no ambiguity in the context. For a vertex set $V'\subseteq V$, the induced subgraph $G[V']$ is the graph with vertices set $V'$ and edge set $\{(u,v)\mid u,v\in V',(u,v)\in E\}$. Further, We define the boundary of a subset of vertices as 
	\[\partial(V')=\{v\mid\exists(u,v)\in E,u\in V',v\not\in V'\}\]
	Moreover, we define $\partial^+(V')=\partial(V')\cup V'$. Two vertex sets $V_1,V_2$ are call connected if $\partial^+(V_1)\cap\partial^+(V_2)\not=\emptyset$.

	A vertex cut is a vertex set $S\subseteq V$ such that $G[V\backslash S]$ is not connected. $|S|$ is called the size of the vertex cut. We also use the 3-tuple $(L,S,R)$ to represent a vertex cut, where $L\cup S\cup R=V$, $L,S,R$ are mutually disjoint, and $\partial(L)\subseteq S$.

	\paragraph{CONGEST model.} Suppose the communication happens in the network $G=(V,E)$. In the CONGEST model, time is divided into discrete time slots, where each slot is called a \emph{round}. Throughout the paper, we always use $n$ to denote the number of vertices in our distributed network, i.e., $|V|$. In each round, each vertex in $V$ can send a $O(\log n)$ bit message to each of its neighbors. At the end of each round, vertices can do arbitrary local computations. A CONGEST algorithm initially specifies the input for each vertex, after several rounds, all vertices terminate and generate output. The time complexity of a CONGEST algorithm is measured by the number of rounds. 
	
	\paragraph{Distributed inputs and outputs.} Since the inputs and outputs to the distributed network should be specified for each vertex, we must be careful when we say something is given as input or is output. Here we make some assumptions. For the network $G=(V,E)$, we say a subset of vertices (or a single vertex) $V'\subseteq V$ is the input or output if every vertex is given the information about whether it is in $V'$. We say a subgraph $H$ (a subset of edges, for example, paths or circles) is the input or output if every vertex knows the edges in $H$ adjacent to it. We say a number is input or output (for example, $\conn$), we normally mean the number is the input or output of every vertex unless otherwise specified.
	
	\paragraph{Dilation and congestion} For $k$ independent CONGEST algorithms $\mathcal{A}=\{A_i\mid i\in[k]\}$ on the same network $G=(V,E)$, the dilation of $\mathcal{A}$ is defined to be $d_{\mathcal{A}}=\max_{i\in[k]}d_i$ where $d_i$ is the round complexity of algorithm $A_i$, and the congestion of $\mathcal{A}$ is defined to be $c_{\mathcal{A}}=\max_{e\in E}\left(\sum_{i\in[k]}c^e_i\right)$, where $c^e_i$ is the number of messages sent through edge $e$ by algorithm $A_i$. The following lemma is taken from \cite{ghaffari2015near}. We will frequently use this lemma in our algorithm description. 
	
	\begin{lemma}\label{lem:randomdelay}
		All algorithm in $\mathcal{A}$ can be simulated in $\tOh(d_{\mathcal{A}}+c_{\mathcal{A}})$ rounds.
	\end{lemma}
	
	As an example, consider growing BFS trees with depth $d$ starting from $c$ vertices, one can see that this can be done in $\tOh(d+c)$ rounds.
	
	\subsection{Vertex Residual Graph}\label{subsec:vertexresidualgraph}
	
	In this section, we will define \emph{vertex residual graph}. We will define a directed graph $G'$, such that finding a directed path on $G'$ is equivalent to finding a $(G,s,P)$-Augmenting Path defined in \Cref{def:overviewaugmentingpath}.
	
	We use ideas from the well-known reduction from vertex connectivity to edge connectivity. We split each vertex $v$ into two vertices $v^{in},v^{out}$, with a directed edge from $v^{in}$ to $v^{out}$. For each edge $(u,v)$ in the original graph, we build an edge from $u^{out}$ to $v^{in}$. Moreover, the residual graph is the graph reversing edge directions on $P$, i.e., for each edge $(u,v)\in E(P)$ we reverse the edge direction of edge $(u^{out},v^{in})$ and for any $u\in V(P)$ we reverse the edge direction of edge $(u^{in},u^{out})$. Since only edges with one edge vertex in $V_I(P)$ will change direction, in the following definition, we only duplicate vertex $v\in V_I(P)$ to $v^{in},v^{out}$, vertex not in $V_I(P)$ can combine $v^{in},v^{out}$ into just one vertex $v^{out}$. 
	
	\begin{definition}[Vertex residual graph]\label{def:augmentingPath}
		Given an undirected graph $G=(V,E)$, a vertex $s\in V$ and a set $P$ containing $k$ internally vertex disjoint simple paths starting from $s$, we define the \textbf{vertex residual graph} on $(G,P)$ as the directed graph $G'=(V',E')$. Let $X=V_I(P)$ be all the internal vertices of $P$.
		\[V'=\{u^{out}\mid u\not\in X\}\cup\{u^{in},u^{out}\mid u\in X\}\]
		\begin{equation*}
			\begin{aligned}
				E'= & \{(u^{out},v^{out})\mid v\not\in X,(u,v)\in E\}\cup\{(u^{out},v^{in})\mid v\in X,(u,v)\in E,(u,v)\not\in E(P),(v,u)\not\in E(P)\}\\
				&\cup\{(v^{in},u^{out})\mid v,u\in X,(u,v)\in E(P)\}\cup\{(v^{out},u^{out})\mid (u,v)\in E(P),v\not\in X\}\\
				& \cup\{(u^{out},u^{in})\mid u\in X\}\\
			\end{aligned}
		\end{equation*}
		We say $s\in G$ is the projection of vertex $s^{in},s^{out}\in G'$, denoted as $M(s^{in})=M(s^{out})=s$. Similarly, for a path $p'$ in $G'$, $M(p')$ is the path on $G$ that maps each vertex in $p'$ to its projection, maintaining the order, and combining consecutive repeated vertices mapped from $(u^{out},u^{in})$. We call a vertex in $V'$ as \emph{in-vertex} or \emph{out-vertex} according to its superscript, i.e., whether it is $v^{in}$ or $v^{out}$ for some $v\in V$. 
	\end{definition}
	
	One should keep in mind the following relationship between \Cref{def:overviewaugmentingpath} and \Cref{def:augmentingPath}: if a path reaches $v^{in}$, that means the path enters $v$ from a vertex other than $suc_p(v)$ and must go to $pre_p(v)$. 
	
	The following lemmas show how to relate finding paths in a vertex residual graph to increasing the number of internally vertex disjoint paths. For an edge $(u,v)$, the reversed edge is defined to be $(u,v)^T=(v,u)$. For an edge set $E$, the reversed edge set is $E^T=\{(v,u)\mid (u,v)\in E$\}. The symmetric difference of $E$ is defined to be $\oplus(E)=E-E\cap E^T$. 
	i.e., all the converse edges are cancelled.
	\begin{lemma}[Augmenting; Proof in Appendix~\ref{app:residualGraph}]\label{lem:augmentingConnectivity}
		Given an undirected graph $G=(V,E)$, a vertex $s\in V$ and a set $P$ containing $k$ internally vertex disjoint simple paths starting from $s$ and ending at multiset $T\subseteq V$, let $G'$ be the vertex residual graph on $(G,P)$, if there exists a simple directed path $p'$ in $G'$ starting from $s$ ending at $v^{out}$, while the internal vertices of $p'$ do not contain any $t^{out}$ for $t\in T$, then there exists $k+1$ internally vertex disjoint simple paths $P'$ starting from $u$ and ending at multiset $T\cup\{v\}$. 
		
		Moreover, in the subgraph $\oplus(E(P)\cup E(M(p'))$, the maximal connected components containing $s$ is $E(P')$; other maximal connected components of $\oplus(E(P)\cup E(M(p'))$ are vertex disjoint circles.
	\end{lemma}
	
	The following lemma shows how to find a cut if an augmenting path cannot be found. 
	
	\begin{lemma}[Find cut]\label{lem:correspondingCut}
		Given an undirected graph $G=(V,E)$, a vertex $s\in V$ and a set $P$ containing $k$ internally vertex disjoint simple paths starting from $s$. Suppose $G'=(V',E')$ is the vertex residual graph on $(G,P)$, and $S'\subseteq V'$ are all the vertices $s^{out}$ can reach in $G'$. Then let $S=\{v\in V\mid v^{in}\in S',v^{out}\not\in S'\}$, $S$ has size at most $k$ and is a vertex cut in $G$ if $V'\backslash S'\not=\emptyset$.
	\end{lemma}
	
	\begin{proof}[Proof of Lemma~\ref{lem:correspondingCut}]
		First, notice that $S$ can contain at most $1$ vertex in each path $p\in P$. If it contains two vertices $v_i,v_j$ where $p=(....,v_i,....,v_j,...)$, then we know $v^{in}_j\in S'$, which means $s^{out}$ can reach $v^{in}_j$, and $v^{in}_j$ can reach $v^{out}_i$ through the backwards direction of $p$. That lead to a contradiction as $v^{out}_i\not\in S'$. Thus, we have $|S|\le k$.
		
		Denote $L=\{v\in V\mid v^{out}\in S'\}$. We will prove that $\partial(L)=S$ and $V\backslash \partial^+(L)\not=\emptyset$, which means $S$ is a vertex cut.
		
		We first prove that $\partial(L)=S$. To prove $\partial(L)\subseteq S$, suppose $v$ is a neighbor of $u\in L$ where $v\not\in L$, we will prove $v\in S$. Notice that $u^{out}\in S'$ according to the definition of $L$. Therefore, if $v^{in}$ do not exists, then $(u^{out},v^{out})$ is an edge and that contradict the fact that $v^{out}\not\in S'$. Thus, $v^{in}$ exists and $(u^{out},v^{in})$ is an edge, which means $v^{in}\in S'$. Since $v^{out}\not\in S'$, we have $v\in L$. To prove $S\subseteq \partial(L)$, for a vertex $v$ with $v^{in}\in S'$ and $v^{out}\not\in S'$, we first have $v\not\in L$. Since $v^{in}\in S'$, there exists a path $(s^{out},....,u^{out},v^{in})$ (since the edge that go into $v^{in}$ must from an out vertex). We have $u^{out}\in S',u\in L$ and $(u,v)\in E$, which proves $v\in \partial(L)$. 
		
		Then we prove that there exists a destination $t$ of a path $p\in P$ such that $t^{out}\not\in S'$. Otherwise, if $S'$ contains all destinations of $p\in P$, then $S'$ contains all $v^{in},v^{out}\in V(P)$, which means $S'$ contains all the nodes in $V'$ as the graph is connected. Now we get $t\not\in L$, and we also know that $t\not\in S$ according to the definition. Since we proved $\partial(L)=S$, we get $t\in V\backslash \partial^+(L)$.
		
	\end{proof}

	\section{IsolatingSmallCut (Proof of~\Cref{lem:smallAlphaAlgorithm})}\label{sec:smallAlpha}
	In this section, we prove Lemma~\ref{lem:smallAlphaAlgorithm}. We will give details of the algorithm described in \Cref{subsubsec:smallalpha}, in the context of vertex residual graph.
	
	\subsection{Distributed Algorithm Details}
	
	We first described the detailed DFS subroutine for one vertex $u$. Recall that $u$ initially gets an empty flow-path set $P$, and at each loop, it uses DFS to increase the size of $P$ by $1$, by finding an augmenting path in the residual graph. We use $c_0$ to denote a sufficiently large constant.
	
	\begin{algorithm}[H]
		\caption{Augmenting($G,u,P,\alpha,\kappa$)}\label{alg:augmenting}
		\KwData{An undirected graph $G=(V,E)$, a vertex $u\in V$, a set of internally vertex disjoint paths $P$ starting from $u$, integers $\alpha<\sqrt{n},\kappa<n^\frac{1}{4}$.}
		\KwResult{A valid $|P|$-vertex cut, or a set of internally vertex disjoint paths $P'$ starting from $u$, with size $|P|+1$.}
		
		Let $G'=(V',E')$ be the vertex residual graph on $(G,P)$. In the following, we describe the CONGEST algorithm on $G'$. One round in $G'$ can be simulated in two rounds in $G$, one round for in-vertex and one round for out-vertex\;
		
		Let $S'\leftarrow \emptyset$. Before the first round, $u^{out}$ receives a DFS-token from a virtual vertex $u_0$\;
		
		\For{each of the following $c_0\kappa\alpha$ rounds, each vertex $v\not=u_0$\label{line:DFS}}{
			\If{$v\not\in S'$ and received the DFS-token from $w$}{Add $v$ to $S'$ and set $f_v=w$}
			\If{$v\in S'$ and owns a DFS-token}{
				For all edges $(v,v')\in E'$, $v$ communicate with $v'$ to check whether $v'\in S'$\;
				\If{$v'\in S$ for all $v'$}{
					Send the DFS-token back to $f_v$\;
				}
				\Else{
					Send the DFS-token to an arbitrary $v'\not\in S'$\;
				}
			} 
		}
		\If{The DFS-token went back to $u_0$}{
			Use~\Cref{lem:correspondingCut} to output a vertex cut. i.e., $\{v\in V\mid v^{in}\in S',v^{out}\not\in S'\}$\; \label{line:findcut}
		}
		\Else{
			Pick a uniformly random out-vertex $v^{out}\in S'$, find a directed simple path $p'$ from $u^{out}$ to $v^{out}$ and use~\Cref{lem:augmentingConnectivity} to update $P$ to $P'$ (See~\Cref{rem:update})\; \label{line:update}
		}
	\end{algorithm}
	
	\begin{remark}\label{rem:update}
		
		More details about line~\ref{line:update}: Let $H$ be the subgraph containing all edges in $G'$ that has transferred the DFS-token. Clearly, $H$ is connected and contains all paths from $u^{out}$ to vertices in $S'$. To sample $v^{out}$, we sample a random rank in $[1,n^c]$ for every out-vertex and pick the highest one by communicating inside $H$; to find $p'$, we simply follow DFS-token back forward path, which is also inside $H$. To update $P$ to $P'$ using $p'$, recall~\Cref{lem:augmentingConnectivity}. We first map $p'$ to $M(p')$. Then we truncate $M(p')$ at the vertex $t$ fit into one of the following two cases
		\begin{enumerate}
			\item $t$ is an end vertex of a path $p\in P$.
			\item $t\not=u,t\in A$ (for the definition of $A$, see Algorithm~\ref{alg:isolating}).
		\end{enumerate}
		
		we compute $H'=\oplus(E(P)\cup E(trancated[M(p')])\subseteq H$, where $trancated[M(p')]$ is the path from $u$ to $t$ in $M(p')$, and find the connected component of $H'$ containing $u$, divide it into $|P|+1$ internally vertex disjoint paths. 
		
	\end{remark}
	\begin{fact}\label{fac:augmenting}
		Algorithm~\ref{alg:augmenting} has the following properties.
		\begin{enumerate}
			\item It either outputs a valid $|P|$-vertex cut, or internally vertex disjoint paths $P'$.
			\item It has constant congestion inside $H$, and has dilation $O(\kappa\alpha)$. 
			\item When the algorithm ends, $S'$ has size $\Omega(\kappa\alpha)$.
		\end{enumerate}  
	\end{fact}

	The first fact is due to~\Cref{lem:augmentingConnectivity,lem:correspondingCut}. This second fact is straightforward from the algorithm description. The third fact is due to the DFS procedure: each round of the for loop (line~\ref{line:DFS}) either add a new vertex to $S'$, or send back the DFS-token from a vertex in $S'$, while each vertex can send back DFS-token only once. 
	
	The main algorithm will run Algorithm~\ref{alg:augmenting} for all vertices in $A$ simultaneously, which might lead to higher congestion. To avoid congestion, we need the following lemma, which is the restatement of \Cref{lem:overviewSolvecongestion} in the context of the vertex residual graph. 
	
	\begin{lemma}[Path-handshaking; proof in Section~\ref{subsec:tracefollowing}]\label{lem:solvecongestion}
		Let $G=(V,E)$ be an undirected graph and $s_1,s_2,t\in V$. For $i\in\{1,2\}$, suppose $P_i$ is a flow-path set of $s_i$. Let $G'_i$ be the vertex residual graph on $(G,P_i)$, and suppose there exists a simple path $p'_i$ on $G'_i$ starting from $s^{out}_i$, ending at $t^{out}$. Then there exists a path $p'$ with $E(M(p'))\subseteq E\left(\{M(p'_1),M(p'_2)\}\cup P_1\cup P_2\right)$, such that $p'$ is either on $G'_1$ starting from $s^{out}_1$ ending at $s_2^{in}$ or $s_2^{out}$, or on $G'_2$ starting from $s^{out}_2$ ending at $s_1^{in}$ or $s_1^{out}$. 
	\end{lemma}
	
	See Algorithm~\ref{alg:isolating} for our main algorithm in this section. For convenience, we duplicate each edge $(u,v)$ in the communication graph into two parallel edges, one edge transfers message sending from $u^{out}$ and one edge transfers message sending from $u^{in}$. 
	
	\begin{algorithm}[H]
		\caption{IsolatingSmallCut($G,A,\conn,\alpha$)}\label{alg:isolating}
		\KwData{An Undirected graph $G=(V,E)$, a vertex set $A\subseteq V$, integers $\alpha<\sqrt{n},\conn<n^{\frac{1}{4}}$. }
		\KwResult{A valid $\kappa$-vertex cut, or $\bot$.}
		
		Let $P_s=\emptyset$ for any $s\in A$\;
		\For{$i$ from $1$ to $k$\label{line:outerloop}}{
			Let $A'\leftarrow A$\;
			\While{$A'\not=\emptyset$\label{line:while}}{
				Let $Pairs\leftarrow\emptyset$\;
				
				Every vertex $s\in A'$ run Augmenting($G,s,P_s,\alpha,\kappa$) (see Algorithm~\ref{alg:augmenting}). Denote the algorithm as $Alg_s$\;
				
				\For{each of the following $c_0\kappa\alpha$ rounds, each edge $(u,v)$\label{line:forloop}}
				{
					\If{There exist more than two vertices $s$ that $Alg_s$ want to send messages from $u$ to $v$}{
						Suppose those vertices are $v_1,v_2,...,v_\ell$, add pairs $(v_1,v_2),(v_3,v_4),...,(v_{\ell-1},v_{\ell})$ to $Pairs$, and stop $Alg_{v_i}$ for $i\in[\ell]$ (this case is when $\ell$ is even). If there are an odd number of vertices, the last unpaired vertex continues its Augmenting procedure\;\label{line:pairup}
					}
				}
			
				If any non-stopped $Alg_s$ outputs a cut, then the algorithm returns the cut, otherwise $Alg_s$ finishes by updating $P_s$\;\label{line:successAlg}
				
				\ForEach{Pair $(u,v)$ in $Pairs$}{
					Let $H_{(u,v)}$ contain all edges that have transferred message for $Alg_u$ or $Alg_v$. Inside $H_{(u,v)}\cup E(P_u)\cup E(P_v)$, $u^{out}$ (and $v^{out}$) use DFS to find a path to $v^{out}$ (and $u^{out}$). If a path is successfully found, use this path to update $P_u$ (or $P_v$), see~\Cref{rem:update}\; \label{line:pair}
				}
			
				Delete all vertices $s$ who successfully updated $P_s$ (either by finishing $Alg_s$, or by line~\ref{line:pair}) from $A'$\;
			}
		}
	If no cut is output, output $\bot$\;
	\end{algorithm}
	
	Line~\ref{line:pairup} refers to the DFS procedure in Algorithm~\ref{alg:augmenting} loop~\ref{line:DFS}. Line~\ref{line:successAlg} refers to line~\ref{line:findcut},~\ref{line:update} in Algorithm~\ref{alg:augmenting}, running simultaneously for every $Alg_s$ using~\Cref{lem:randomdelay}. Line~\ref{line:pair} is also running simultaneously for every pair in $(u,v)$ using~\Cref{lem:randomdelay}. 
	
	\subsection{Proof of Lemma~\ref{lem:solvecongestion}}\label{subsec:tracefollowing}
		In this section, we will prove the important path-handshaking lemma. Suppose $P_1,P_2$ end at multiset $T_1,T_2$, respectively. According to \Cref{lem:augmentingConnectivity}, there exists internally vertex disjoint paths $P'_1,P'_2$ ending at $T_1\cup\{t\},T_2\cup\{t\}$, where $E(P'_1)\subseteq \oplus(E(P_1\cup \{M(p'_1)\})),E(P'_2)\subseteq \oplus(E(P_2\cup \{M(p'_2))\})$. Denote the edge set $E(P_1\cup P_2\cup \{M(p'_1),M(p'_2)\})$ as a subgraph $H=(V_H,E_H)$. We have $E(P'_1)\cup E(P'_2)\subseteq H$ according to \Cref{lem:augmentingConnectivity}. Let $H'_1,H'_2$ be the vertex residual graph restricted on $H$ (i.e., the subgraph of $G'_i$ that contains edges $(u,v)$ with $M(u)=M(v)$ or $(M(u),M(v))\in H$). Let $S_i$ denote the set of all the vertices $v$ such that $s_i$ can reach $v^{out}$ in $H'_i$. Since $p'_1,p'_2$ both end at $t^{out}$, we have $t\in S_1\cap S_2$. Now we suppose $s^{out}_1$ cannot reach $s^{in}_2,s^{out}_2$on $H'_1$, and $s^{out}_2$ cannot reach $s^{in}_1,s^{out}_1$ on $H'_2$, and try to get contradiction. We first show some properties of $S_i$ and $P'_i$ for $i\in\{1,2\}$.
		\begin{claim}\label{cla:edge}
			If $(u,v)\in \oplus(E(P_i\cup \{M(p'_i)\}))$ and $v^{in}$ or $v^{out}$ is reachable from $s_i^{out}$ in $H'_i$, then $u\in S_i$.
		\end{claim}
		\begin{proof}
			According to the definition of $\oplus$, $(u,v)$ are either in $E(p^*)$ for some $p^*\in P_i$, or in $E(M(p'_i))$. In the former case, $v^{out}$ can reach $v^{in}$, $v^{in}$ can reach $u^{out}$, which means $u\in S_i$. Now suppose $(u,v)\not \in E(P_i)$. In the latter case, we also have $(v,u)\not\in E(P_i)$ according to the definition of $\oplus$. Thus, $u^{in}$ has no edge to $v^{in}$ or $v^{out}$. Since $(u,v)\in E(M(p'_i))$, the only possibility is $u^{out}\in V(p'_i)$. Recall that $p'_i$ is a path from $s_i^{out}$ in $H'_i$, we have $u\in S_i$. 
		\end{proof}
		\begin{claim}\label{cla:failvertex}
			For each path $p\in P'_i$, exactly a prefix of the path is in $S_i$; i.e., either $V(p)\subseteq S_i$, or there exists a vertex $F(p)\in V(p),F(p)\not\in S_i$, such that $u\in S_i$ for any $u\prec_pF(p)$. 
		\end{claim}
		\begin{proof}
			Suppose $p\in P'_i,p=(s_i=v_0,v_1,...,v_{\ell})$. We have $s_i\in S_i$. Suppose $v_m\in S_i$ for some $m>0$, which means $v_m^{out}$ is reachable from $s_i^{out}$ in $H'_i$. Since we have $(v_{m-1},v_m)\in\oplus(E(P_i\cup \{M(p'_i)\}))$, according to \Cref{cla:edge}, we get $v_{m-1}\in S_i$. Thus, for any $\forall u\prec_p v_m$, we have $u\in S_i$, which leads to the claim.
		\end{proof}
		
		\begin{claim}\label{cla:neighbor}
			In $H$, all neighbors of $w\in S_i$ is either in $S_i$ or in $F(p)$ for some $p\in P'_i$. 
		\end{claim}
		\begin{proof}
			Suppose $w'$ is a neighbor of $w$, if $w'\not\in V_I(P_i)$, then there is an edge from $w^{out}$ to $w'^{out}$, which means $w'^{out}$ can be reached from $w^{out}$ in $H'_i$. Thus, $w'\in S_i$. Now suppose $w'\in V_I(P_i)$ and $w'\not\in S_i$. There is an edge from $w^{out}$ to $w'^{in}$, which means $w'^{in}$ is reachable from $s_i^{out}$ in $H'_i$. Suppose $w'\in V_I(p^*)$ for some $p^*\in P_i$, and $p^*=(s_i=v_0,v_1,...,v_{m-1},v_m=w',v_{m+1},...)$, i.e., $v_{m-1},v_{m+1}$ are two vertices adjacent to $w'$ in $p^*$. We first show that $(v_m,v_{m+1})\in\oplus E(P_i\cup \{M(p'_i)\})$. To prove this, we only need to show that $(v_{m+1},v_m)\not\in E(M(p'_i))$. Actually, either $v^{out}_{m+1}$ or $v^{in}_{m+1}$ is in $E(p'_i)$, both means $v^{out}_m$ is reachable from $s_i^{out}$ and $v_m\in S_i$, which is a contradiction. Thus, $(v_{m+1},v_m)\not\in E(M(p'_i))$ is true. Then we show that $v_m\in V(P'_i)$. To prove this, notice that $(v_m,v_{m+1})$ is either in a circle, or in $E(P'_i)$ according to \Cref{lem:augmentingConnectivity}. Suppose $(v_m,v_{m+1})$ is in a circle $C$. We have $C\subseteq\oplus E(P_i\cup \{M(p'_i)\})$ and $s_i\not\in C$. Since $P_i$ are internally vertex disjoint simple paths, there must exist an edge in $C$ that is in $E(M(p'_i))$. An end point of this edge must be in $S_i$. According \Cref{cla:edge}, all vertices on the circle are in $S_i$, which is a contradiction to $v_m\not\in S_i$. Thus, the only possibility is $(v_m,v_{m+1})\in E(P'_i)$. Let $p''\in P'_i$ be the path that $v_m=w'\in V(p'')$, and $(w'_{pre},w')\in E(p'')$. Recall that $w'^{in}$ is reachable from $s_i^{out}$ in $H'_i$. According to \Cref{cla:edge}, $w'_{pre}\in S_i$. According to \Cref{cla:failvertex}, $w'=F(p'')$, which finish the proof. 
		\end{proof}
		
		We divide the paths in $P_1\cup P_2$ into three types, defined as follows. For $i=1,2$, we define $i'=2,1$ as the opposite index. Suppose path $p$ starts at $s_i$ and ends at $t_p$, $p$ is called
		\begin{enumerate}
			\item \emph{{type 1}}, if $F(p)$ do not exists (which means $V(p)\subseteq S_i$ according to \Cref{cla:failvertex}), and $t_p\in S_{i'}$.
			\item \emph{{type 2}}, if $F(p)$ exists, and $F(p)\in S_{i'}$.
			\item \emph{{type 3}}, if it is not type 1 or type 2 path.
		\end{enumerate}
		Among all the paths, let the number of {type 1}, {type 2}, {type 3} paths be $n_1,n_2,n_3$ respectively. For a type 1 or type 2 path $p=(s_i=w_0,w_1,w_2,...,w_\ell,...)$, where we set $w_{\ell}=t_p$ if $F(p)$ do not exists, and $w_{\ell}=F(p)$ if $F(p)$ exists. we define $v(p)=w_m$ as the vertex with the smallest index $m\in(0,\ell)$ satisfying $w_{m+1}\in S_{i'}$ and $w_m\not\in S_{i'}$. Since we have $w_{\ell}\in S_{i'}$ and $w_1\not\in S_{i'}$ (If $w_1\in S_{i'}$, then $s_{i'}$ can reach $w_1^{out}$, in which case $s_j$ can reach $w_0^{in}$ or $w_0^{out}$, but remember that we assume $s_j$ cannot reach $w_0^{in}$ or $w_0^{out}$), $v(p)$ must exists. We first show that for a different type 1 or type 2 path $p'\not=p$, it must hold that $v(p)\not=v(p')$. Otherwise, suppose $v(p)=v(p')$, where $p'=(s_j=w'_0,w'_1,...,w'_{m'}=w_{m}=v(p)=v(p'),w'_{m+1},...)$. Note that $i\not=j$ since $P'_i$ are internally vertex disjoint paths. Further, $w_{m}\in S_j$, according to \Cref{cla:failvertex}. That leads to a contradiction to the definition of $v(p)$. 
		
		Since there are $n_1+n_2$ type 1 or type 2 paths, we get in total $n_1+n_2$ such vertices $v(p)$ and corresponding edge $(w_{m}=v(p),w_{m+1})$, where $w_{m+1}$ is on $S_{i'}$ and $w_{m}$ is not in $S_{i'}$. Thus, according to \Cref{cla:neighbor}, $w_{m-1}\in F(p^*)$ for some $p^*\in P'_{i'}$. Path $p^*$ is an {type 2} path since $F(p^*)\in S_i$. There are $n_2$ {type 2} paths, and $v(p)=w_{m}$ is distinct for different type 1 or 2 paths $p$, we have $n_1+n_2=n_2$, which means $n_1=0$. However, remember that $P'_1$ contains a path ending at $t$, and $t\in S_1\cap S_2$, which is a type 1 path. That leads to a contradiction since type $1$ paths do exist. 
		
		\subsection{Analysis of Algorithm}\label{subsec:proofofloweralpha}
			\paragraph{Round complexity.} We first bound the number of while loops in line~\ref{line:while}. 
			
			\begin{lemma}\label{lem:numberofwhileloop}
				For any $i$, the while loop in line~\ref{line:while} contains $O(\log n)$ loops.
			\end{lemma}
			\begin{proof}
				We will prove that each while loop will decrease the size of $A'$ by at least by half. Since pairs in $Pairs$ are disjoint, and vertices not inside $Pairs$ are all deleted from $A'$, we just need to prove line~\ref{line:pair} will find a path either from $u^{out}$ to $v^{out}$ or from $v^{out}$ to $u^{out}$. Notice that $(u,v)$ are inside $Pairs$ because in line~\ref{line:pairup}, the algorithm $Alg_u,Alg_v$ collide at some edge $(w_1,w_2)$. Remember that we duplicate each edge into two edges one for transferring messages from out-vertex and one for transferring messages from in-vertex, so there are two cases to consider
				\begin{itemize}
					\item $Alg_u,Alg_v$ both send from out-vertex $w^{out}_1$. That means the DFS-token from $u^{out}$ and $v^{out}$ both arrives $w^{out}_1$, which means there is a path from $u^{out}$ and $v^{out}$ to $w^{out}_1$ in the residual graphs, satisfying the precondition of~\Cref{lem:solvecongestion}. Also notice that the mapping of these paths is included in $H_{(u,v)}$. Thus, by DFS searching in $H_{(u,v)}\cup E(P_u)\cup E(P_v)$, at least one of $u^{out}$ or $v^{out}$ will find path to $v^{in},v^{out}$ or $u^{in},u^{out}$. Let us consider the case where $u^{out}$ reach $v^{in},v^{out}$, we will argue that $v^{in}$ does not exist in the residual graph of $u$: that is because the truncating of~\ref{rem:update}, any internal vertex of a path in $P_u$ cannot contain $v$. Finally, we have $u^{out}$ reach $v^{out}$ in the residual graph. 
					
					\item $Alg_u,Alg_v$ both send from in-vertex $w^{in}_1$. Recall that in~\Cref{def:augmentingPath}, $w^{in}_1$ has only one out-neighbor, which is an out-vertex $w^{out}_2$. Besides, the edge $(w_2,w_1)$ is in both $P_u,P_v$. Therefore, $u^{out},v^{out}$ both have paths to $w^{out}_2$ in the residual graphs, and the mapping of these paths are inside $H_{(u,v)}\cup E(P_u)\cup E(P_v)$. Thus, by DFS searching in $H_{(u,v)}\cup E(P_u)\cup E(P_v)$, at least one of $u$ or $v$ will successfully update $P_u$ or $P_v$ based on the same argument as above.
				\end{itemize}
				
				 This finishes the proof. 
			\end{proof}
		
			Then we bound the complexity of line~\ref{line:successAlg},~\ref{line:pair}.
			\begin{lemma}\label{lem:congestion1}
				For each $v\in A'$ such that $Alg_v$ is not stopped, let $H_v$ be a subgraph containing all edges involved in $Alg_v$. $H_v$ has dilation $O(\kappa\alpha)$, and for any edge $e$, the number of different $s\in A'$ where $e\in H_{s}$ is bounded by $O(\kappa\alpha)$. 
				
				For each pair $(u,v)$ in $Pairs$, let $H'_{(u,v)}=H_{(u,v)}\cup E(P_u)\cup E(P_v)$ in line~\ref{line:pair}, the dilation of $H'_{(u,v)}$ is bounded by $\tOh(\kappa^2\alpha)$, and for any edge $e$, the number of different pairs $(x,y)$ where $e\in H_{(x,y)}$ is bounded by $\tOh(\kappa^2\alpha)$. 
			\end{lemma}
			\begin{proof}
				According to~\Cref{fac:augmenting}, $H_v$ has dilation $O(\kappa\alpha)$. Consider the for loop~\ref{line:forloop}, each round each edge can transfer one message, which means each edge can be included in at most one $H_v$ in each round. Since there are $O(\kappa\alpha)$ rounds, the first part of the lemma is proved. 
				
				The above arguments also hold for $H_{(u,v)}$. To bound the dilation and congestion for $H'_{(u,v)}$, we focus on analysing $E(P_u)$ and $E(P_v)$. The updating rule of $P_u$ guarantees that if an edge is included in $P_u$, it must be transferring message in line~\ref{line:forloop} for $Alg_u$ at some previous loops. Since in each round, each edge can transfer one message, and line~\ref{line:forloop} runs in $O(\kappa\alpha)\cdot O(\log n)\cdot \kappa=O(\kappa^2\alpha)$, the lemma is proved. The term $O(\log n)$ comes from the number of while loops~\ref{line:while} according to~\Cref{lem:numberofwhileloop}, and $\kappa$ term comes from the number of outer loops~\ref{line:outerloop}.
			\end{proof}
			
			The round complexity claimed in~\Cref{lem:smallAlphaAlgorithm} is $\kappa\cdot O(\log n)\cdot \tOh(\kappa^2\alpha)=\tOh(\kappa^3\alpha)$, the third term is the complexity of inner loop and the first two terms are the number of outer loops and inner loops.
			
			\paragraph{Correctness.} Then we prove the correctness. We first prove that, if a cut is output, it must be a valid cut. Recall in Algorithm~\ref{alg:augmenting}, a cut is output iff. the token goes back to $u_0$. This can only happen when $S'$ contains all the vertices that $u^{out}$ can reach. According to~\Cref{lem:correspondingCut}, $\{v\in V\mid v^{in}\in S',v^{out}\not\in S'\}$ is a cut as long as $V'\backslash S'\not=\emptyset$. The latter claim is because $\kappa<n^{1/4},\alpha<\sqrt{n}$. As $|S'|=O(\kappa\alpha)$, $V'\backslash S'$ must be non-empty. 
			
			Now suppose there exists $s\in A,L\subseteq V$ such that $|\partial(L)|<\conn,A\cap \partial^+_v(L)=\{s\},|L|\le\alpha$, we will prove that $\bot$ will be output with at most constant probability. Denote $R=V\backslash\partial^+(L)$. If all paths in $P_s$ end at $R$, then according to Lemma~\ref{lem:augmentingConnectivity}, there exists $\conn$ internally vertex disjoint path between $s$ and $R$, which contradiction the fact that $|\partial(L)|<\conn$. Let $i$ be the first outer loop (line~\ref{line:outerloop}) where the ending set of $P_s$ contains a vertex not in $R$. Let $p'$ be the path found in line \ref{line:update}. If $p'$ is truncated by~\Cref{rem:update} by $t$ where $t$ is an end vertex of a path $p\in P$, the ending set of $P_s$ is the same in the $(i-1)$-th loop; if $p'$ is truncated by~\Cref{rem:update} by $t$ where $t\not=s,t\in A$, then $t$ must be in $R$. Thus, in the $i$-th loop, $p'$ must end at a vertex in $\partial^+(L)$. 
			
			Let $I_i$ be the event that the first outer loop (line~\ref{line:outerloop}) where the ending set of $P_s$ contains a vertex not in $R$ is loop $i$. Let $p'$ be the path found in line \ref{line:update} in loop $i$. The endpoint of $p'$ is a random vertex among $\Omega(\conn\alpha)$ vertices, which lies in $\partial^+(L)$ with probability bounded by $O(1/\conn)$ since $|\partial^+(L)|=O(\alpha+\conn)$. Thus, the event $I_i$ happens with probability $O(1/\conn)$. If the algorithm return $\bot$, then one of $I_1,I_2,...,I_\kappa$ must happen. By union bound, "the algorithm return nothing" has probability bounded by $O(1/\conn)\cdot\conn$. By letting the constant hidden in $O$ sufficiently small, the probability is bounded by a constant. 

		\section{SingleSourceLocalCut (Proof of~\Cref{lem:largeAlphaAlgorithm})}\label{sec:largeAlpha}
		In this section we prove Lemma~\ref{lem:largeAlphaAlgorithm}.
		
		\subsection{Path Centered Clustering}\label{subsec:pathcenteredclustering}
		We first give the definition of \emph{paths centered clustering} promised at~\Cref{rem:canaugment}. For a path $p=(v_0,v_1,...,v_\ell)$ and vertices $v_i,v_j$ with $i<j$, we write $v_i\prec_pv_j$ to denote $v_i$ precedes $v_j$ on path $p$.

		\begin{definition}[Paths centered clustering]\label{def:clustering}
			For an undirected graph $G(V,E)$ with diameter $D$ and a set of $k$ simple paths $P$, a \textbf{paths centered clustering} on $(G,P)$ is a tuple $\mathcal{C}=(\mathcal{S},Centers,Rep,LU)$, where $\mathcal{S}$ is a partition\footnote{A set of vertex sets $\mathcal{S}$ is defined to be a \emph{partition} of a vertex set $U$, if any two sets in $\mathcal{S}$ are disjoint, and the union of $\mathcal{S}$ is $U$. } of $V\backslash V(P)$; $Centers,Rep,LU$ are functions on $\mathcal{S}$, each $S\in\mathcal{S}$ is called a cluster. For any $S\in\mathcal{S}$, we have
			\begin{enumerate}
				\item $Centers[S]$ is a vertex set containing at most one vertex in each path $p\in P$, and $Rep[S]\in Centers[S]$. 
				\item $G[S]$ is connected; each vertex in $Centers[S]$ is a neighbor of some vertex in $S$; the subgraph $\{(u,v)\mid u\in S,v\in S\cup Centers[S]\}$ has diameter at most $kD$.
				\item Suppose $Rep[S]$ is on path $p\in P$. We write $S_v$ as the cluster $S_v\in\mathcal{S}$ containing $v$ if $v\in V\backslash V(P)$. Then there exists an edge $(u,v)$ such that $u\in S$, and $v$ satisfies
				\begin{itemize}
					\item If $LU[S]=0$, then either $v\in V(p),v\prec_pRep[S]$, or there exists $c\in Centers[S_v]$ such that $c\prec_pRep[S]$. $S$ is call a \textbf{lower cluster} of $Rep[S]$.
					\item If $LU[S]=1$, then either $v\in V(p),Rep[S]\prec_pv$ or there exists $c\in Centers[S_v]$ such that $Rep[S]\prec_pc$. $S$ is call a \textbf{upper cluster} of $Rep[S]$.
				\end{itemize}
			\end{enumerate}
		\end{definition}
	
\begin{figure}[H]
			\centering
			\centering
			\includegraphics[scale=0.8]{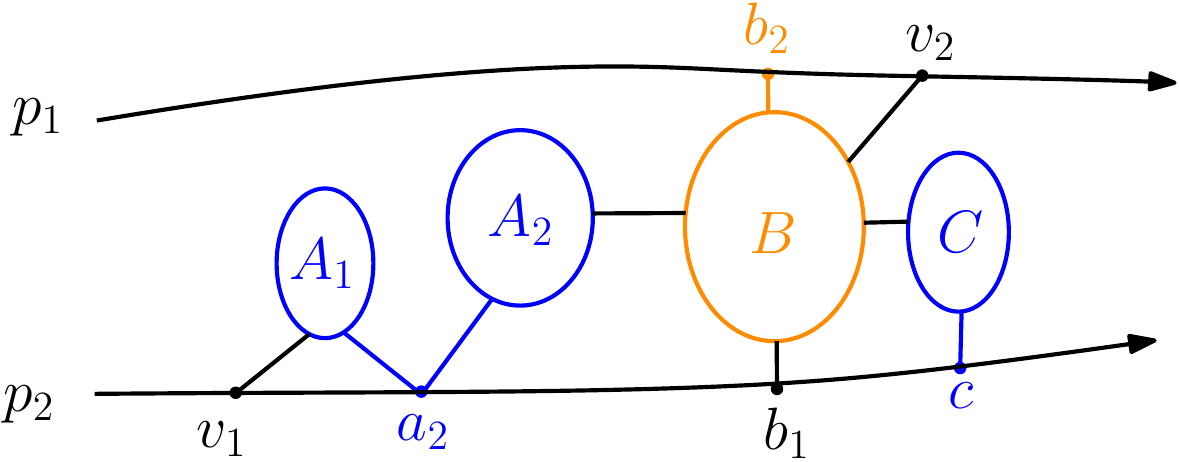}
			\setcaptionwidth{0.85\textwidth}
			\caption{We have $P=\{p_1,p_2\}$, $A_1,A_2,B,C\in\mathcal{S}$, $Centers[A_1]=Centers[A_2]=\{a_2\},Centers[B]=\{b_1,b_2\},Centers[C]=\{c\},Rep[A_1]=Rep[A_2]=a_2,Rep[B]=b_2,Rep[C]=c$, and $LU[A_1]=LU[C]=0,LU[A_2]=LU[B]=1$. $A_1,C$ are lower clusters because they are connected to $v_1,B$ separately; $A_2,B$ are upper clusters because they are connected to $B,v_2$ separately. Moreover, for any vertices $u_1\in A_1,u_2\in A_2,u_3\in C$, we have $u_1\preceq_\mathcal{C}u_2\preceq_{\mathcal{C}}u_3$. Notice that $B$ and $C$ are not comparable with respect to $\preceq_{\mathcal{C}}$, since $Rep[B]$ is on $p_1$ but $Rep[C]$ is on $p_2$.}.\label{fig:clustering}	
		\end{figure}

		Recall that in the simplified version defined in~\Cref{subsubsec:largealpha}, the clustering is a partition of $V$ and each cluster contains exactly one vertex in the flow-paths $P$ as its center. However, it is different in the above definition: clustering is a partition of $V\backslash V(P)$, each cluster $S$ contains no vertex in $P$, but still has a unique "representative" $Rep[S]$ adjacent to $S$ in $P$. Moreover, each node in $P$ might be the "representative" of several different clusters. The function $LU$ defines whether $S$ is connected to the upper part or the lower part of the "representative" path. See~\Cref{fig:clustering} as an example.

		The following definition defines a partial order $\preceq_{\mathcal{C}}$ on $V$ based on a paths centered clustering $\mathcal{C}$. Using the following definition, the problem mentioned in~\Cref{rem:canaugment} is solved, see~\Cref{rem:orderingReachability}.
		
		\begin{definition}[Paths centered order]\label{def:ordering}
			A \textbf{paths centered order} on the paths centered clustering $\mathcal{C}=(\mathcal{S},Centers,Rep,LU)$ on $(G,P)$ is a partial order $\preceq_{\mathcal{C}}$ defined as follows. We first extend functions $Rep,LU$ to all vertices: for $v\in V(P)$, $Rep[v]=v,LU[v]=1$; for $v\not\in V(P)$, let $S_v\in\mathcal{S}$ be the cluster that $v$ is in, then $Rep[v]=Rep[S_v],LU[v]=LU[S_v]$. For $u,v\in V$, we define $u\preceq_{\mathcal{C}}v$ iff. $Rep[u],Rep[v]$ are both on path $p\in P$ and $Rep[u]\prec_{p}Rep[v]$, or $Rep[u]=Rep[v],LU[u]\le LU[v]$. 
		\end{definition}
		
\begin{figure}[H]	
			\centering
			\centering
			\includegraphics[scale=0.8]{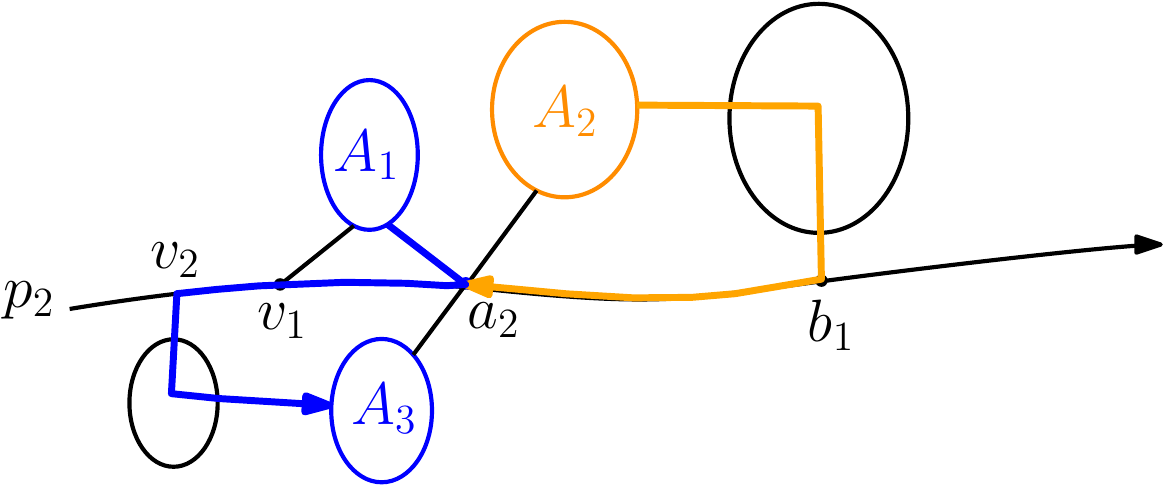}
			\setcaptionwidth{0.85\textwidth}
			\caption{$Rep[A_1]=Rep[A_2]=Rep[A_3]=a_2,LU[A_1]=LU[A_3]=0,LU[A_2]=1$, i.e., ,$A_1,A_3$ are two lower clusters of $a_2$, $A_2$ is an upper cluster of $a_2$. Any vertex in $A_2$ can follow the orange path to reach $a^{out}_2$, then reach any vertices in $A_1,A_2,A_3$. Any vertex in $A_1$ can follow the blue path to reach any lower clusters of $a_2$ (for example, $A_3$).}\label{fig:ordering}	
		\end{figure}

		\begin{remark}\label{rem:orderingReachability}
			The paths centered ordering is defined based on the following intuition: if $u\preceq_{\mathcal{C}}v$, then $u$ can reach $v$ in the vertex residual graph $G'$ on $(G,P)$. See~\Cref{fig:ordering} for an example.
		\end{remark}
		
		
		The following lemma shows there exists a fast algorithm to compute paths centered clustering.
		\begin{lemma}[Clustering; Proof in Section~\ref{subsec:clustering}]\label{lem:clustering}
			On an undirected graph $G=(V,E)$ with diameter $D$, there exists a CONGEST model algorithm given a set of vertex disjoint simple paths $P$, either output a vertex cut with size at most $|P|$, or compute a path centered clustering on $(G,P)$, denoted as $(\mathcal{S},Centers[S],Rep[S])$, where each vertex in $S\in\mathcal{S}$ knows $S,Centers[S],Rep[S],LU[S]$. 
			The algorithm has dilation $\tilde{O}(|P|^2D)$ and congestion $\tOh(|P|^2)$. 
		\end{lemma}
	
		\subsection{Algorithm Overview}\label{subsec:LargeAlphaAlgorithmOverview}
		\paragraph{Partial Virtual Graph.}
		Recall that in \Cref{subsubsec:largealpha}, we showed the framework of reachability in the CONGEST model: construct a virtual graph which depicts the reachability of the original graph. We also showed that it is not efficient to build the virtual graph on all sample hubs. In fact, any edge in the virtual graph starts at \emph{active hubs}. The reason is that if we can reach a non-active hub, then we can find the desired path and can stop. This leads to the following definition of \emph{partial virtual graph}, where the partial virtual graph is only guaranteed to either depict the reachability of the graph or can reach the desired destination.
		
		For two vertices in graph $G$, we use $u\stackrel{G}{\to}v$ to denote an arbitrary path from $u$ to $v$ in $G$. 
		\begin{definition}[Partial Virtual Graph]\label{def:semivirtual graph}
			For a directed graph $G=(V,E)$ and a set $T\subseteq V$, a \textbf{partial virtual graph} on $(G,T)$ with dilation $d$ is a virtual graph $G'=(V',E')$ where $V'\subseteq V$ satisfying the following property: For any two vertices $s\in V',v\in V$, if $s\stackrel{G}{\to}v$ exists, then there exists $v'\in V',s\stackrel{G'}{\to}v'$ and $v'$ satisfies one of the following two conditions.
			\begin{enumerate}
				\item $v'\in T$.
				\item $v'$ has distance at most $d$ to $v$ in $G$.
			\end{enumerate}
		\end{definition}
		
		The following is the schematic of our algorithm. We omit most of the details to give the reader a high-level idea of what the algorithm is doing. The implementation details will be presented in the following sub-sections. 
		\begin{construction}{Schematic of SingleSourceLocalCut($G,s,t,\conn,\alpha$)}
			\begin{itemize}
				\item \textbf{Inputs:} An undirected graph $G=(V,E)$, vertices $s,t\in V$, integers $\conn,\alpha,d\in\mathbb{N}$ with $\conn<n^\frac{1}{4}$.
				\item \textbf{Outputs:} A valid $\kappa$-vertex cut, or $\bot$.
				\item Initially set $P=\emptyset$. $P$ is a set of internally vertex disjoint paths starting from $s$. Let $d=\kappa^{1.5}\sqrt{n}$. Repeat the following steps for $\conn$ loops. 
				\begin{enumerate}
					\item Each vertex becomes a \emph{hub} with probability $p=1/d$. $s,t$ become hubs with probability $1$. Use Lemma~\ref{lem:clustering} on $(G,P)$ to get a path centered clustering $\mathcal{C}$, or \textbf{Return} a vertex cut. Each hub $v$ becomes an \emph{active hub} if the number of hubs $u$ with $u\preceq_{\mathcal{C}}v$ is bounded by $O(\conn\alpha p\log n)$. \label{step1}
					\item Let $G'$ be the vertex residual graph on $(G,P)$. If $v$ is a hub or active hub in $V$, then $v^{out}$ is a hub or active hub in $V'$. Let $H$ contain all the hubs and $H_a$ contain all the active hubs in $V'$. Each vertex in $H_a$ broadcast $\tOh(\conn)$ bits messages to the whole graph (will be specified in \Cref{subsec:buildpartialvirtual graph}, analogue to "Downward edges" and "upward edges" described in~\Cref{subsubsec:largealpha}). Using these informations, a partial virtual graph $G''$ with dilation $\Theta(d\log n)$ on $(G',T=\{t^{out}\}\cup H\backslash H_a)$ can be known by all vertices. \label{step2}
					\item Let $H_r$ contain all the vertices that $s$ can reach in $G''$. if $|H_r|=\Omega(\conn\alpha p\log n)$, then sample a uniformly random $h_1$ from $\Theta(\conn\alpha p\log n)$ vertices in $H_r$, and let $p''=s\stackrel{G''}{\to}h_1$; otherwise, if $t^{out}\in H_r$, let $p''=s\stackrel{G''}{\to}t^{out}$; otherwise, if there exists $h_0\in H_r\cap(H\backslash H_a)$, then sample a uniformly random hub $h^*$ among all hubs $h$ with $h\preceq_\mathcal{C}h_0$ and let $p''=s\stackrel{G''}{\to}h_0\stackrel{G'}{\to}h^*$; otherwise, for each $h\in H_r$, let $S'$ contain all the vertices that a vertex in $H_r$ can reach with distance $\Theta(d\log n)$, if $S'$ has no outgoing edges, \textbf{Return} the vertex cut $\{v\in V\mid v^{in}\in S',v^{out}\not\in S'\}$, otherwise \textbf{Return} $\bot$. \label{step3}
					\item Map $p''$ into a path $p'$ in $G'$. Update $P$ by the augmenting path $p'$ using Lemma~\ref{lem:augmentingConnectivity}. \label{step4}
				\end{enumerate}
			\item If no cut is output, \textbf{Return} $\bot$. 
			\end{itemize}
		\end{construction}
		
		\paragraph{Distributed implementation organization.} One can see that there are four steps in each loop. We will describe the implementation details for each step in the following subsections. 
		
		\begin{description}
			\item[Step \ref{step1} (\Cref{subsec:findhubs}):] We will describe how to get \emph{hubs} and \emph{active hubs}, this can be done easily by aggregation through clusters and paths.
			\item[Step \ref{step2} (\Cref{subsec:buildpartialvirtual graph}):] We will define the partial virtual graph $G''$ by giving $5$ types of edges in $G''$. We will also show that those types of edges can be constructed by broadcasting $\tOh(\conn)$ bits of messages by each active vertices.  
			\item[Step \ref{step3} (\Cref{subsec:findaugmentingpath,subsec:substepfindpathfromh0toh}):] There are four \emph{if-else} possibilities in step \ref{step3}, we will give more details for each possibility in \Cref{subsec:findaugmentingpath}. One important detail is how to get the path $h_0\stackrel{G'}{\to}h^*$, which basically shows how to get the path according to \Cref{rem:orderingReachability}. This is in \Cref{subsec:substepfindpathfromh0toh}.
			\item[Step \ref{step4} (\Cref{subsec:changpathupdateP}):] The final step will give details about mapping paths in $G''$ into paths in $G'$, by distributively mapping each type of edge in $G''$ to paths in $G'$. The final step also contains the details of updating $P$.
		\end{description}

		\subsection{Step \ref{step1}: Find hubs and active hubs.}\label{subsec:findhubs}
		We first introduce a basic algorithm that we will frequently use. The proof is deferred to Appendix~\ref{app:mergeClusters}. The lemma mainly shows how to efficiently aggregate values in a path by divide and conquer. 
		\begin{lemma}[Path aggregation; Proof in Appendix~\ref{app:mergeClusters}]\label{lem:pushlabel}
			There exists a CONGEST algorithm given a directed graph $G=(V,E)$ with undirected diameter $D$, given a path $p=(v_0,v_1,...,v_k)$, an integer $d$ such that $1\le d\le k$, each vertex $v_i$ on the path get a polynomial bounded integer $x[v_i]$ (we treat repeated vertices on the path as different vertices); each vertex $v_i$ outputs the value $\sum_{0\le j\le i}x[v_j]$. The algorithm has dilation $\tilde{O}(d+D)$ and congestion $\tilde{O}(k/d)$.
		\end{lemma}
		
		Now we show how to compute hubs and active hubs. Firstly, we calculate for each vertex $v\in V(P)$, its path's index $PI[v]$ (defined later) and its own index $I[v]$ (defined later) on the path. This can be done by running the algorithm in Lemma~\ref{lem:pushlabel} two times for each path $p_i\in P$. Firstly $x_j=1$ for all $j$ to get $I[v]$; secondly $x_j=0$ for all $j$ except $x_0=i$ to get $PI[v]$. Since there are at most $\conn$ paths, each has length bounded by $\tOh(\conn\alpha)$, this cause a dilation $\tilde{O}(d+D)$ and congestion $\tilde{O}(\conn^2\alpha/d)$. 
		
		Then we use Lemma~\ref{lem:clustering} on $(G,P)$ to create $\mathcal{C}=(S,Centers,Rep,LU)$. Notice that $H_S=(S\cup Centers[S],\{(u,v)\mid u\in S,v\in S\cup Centers[S]\})$ is a subgraph with diameter at most $O(kD)$, and are disjoint for different $S$. Thus, in the following, we always assume the aggregation problem inside $S$ happens in $H_S$, and are computed in parallel for all $S$. Each vertex $v\in S$ can get the information $PI[Rep[S]],I[Rep[S]]$, by broadcasting inside $H_S$. Recall that we sample each vertex in $G$ as \emph{hub} with probability $p=1/d$, and $s$ is a hub. For each $v\in V(P)$, let $Num_L[v]$ denote the number of hubs inside all lower clusters of $v$, and let $Num_R[v]$ denote the number of hubs in all upper clusters of $v$ plus an indicator variable that equals to $1$ if $v$ is a hub. This can be computed by collecting the number of hubs inside $S$ and sending it to $Rep[S]$ using $H_S$. This procedure case dilation $\tOh(\conn D)$ and congestion $\tOh(1)$. 
		
		Recall that we defined $\preceq_{\mathcal{C}}$ as the paths centered ordering in Definition~\ref{def:ordering}. Now we want to compute for each hub $h$, the number of hubs $h'$ such that $h'\preceq_{\mathcal{C}}h$, denoted as $\SNum_{\mathcal{C}}[h]$. This can be done by running the algorithm in Lemma~\ref{lem:pushlabel} for each path $p$, each vertex $v$ on the path receives input value $Num_L[v]+Num_U[v]$. Upon each $v$ receives the output of the algorithm $\sum_{v'\le_pv}(Num_L[v']+Num_U[v'])$, it sends the output to all its upper clusters and sends the output minus $Num_U[v]$ to all its lower clusters. At last each hub $h$ becomes an \emph{active hub} if $\SNum_{\mathcal{C}}[h]=\tOh(\conn\alpha p)$. This guarantees that the number of active hubs is bounded by $\tOh(\conn^2\alpha p)$. This can be done with dilation $\tOh(d+\conn D)$ and congestion $\tOh(\conn^2\alpha/d)$.
		
		\subsection{Step \ref{step2}: Build partial virtual graph.}\label{subsec:buildpartialvirtual graph}
		
		In this step, we show how to build the partial virtual graph $G''$ on the vertex residual graph efficiently, mainly by reducing the number of edges in the partial virtual graph while maintaining the mutual reachability relationship and broadcasting a small amount of information to let all vertices know all edges in $G''$. 
		
		Recall that $G'$ is the vertex residual graph on $(G,P)$, and $v^{out}$ is a hub or active hub in $G'$ iff. $v$ is a hub or active hub in $G$. There are five types of edges in the partial virtual graph. We will define them and also show what information should each active vertex broadcast to build the edges. Recall that in \Cref{subsubsec:largealpha}, we only give two types of edges, namely upwards edges and downwards edges. The 5 types of edges is an extensions of it in order to make it easy to recover a path in $G'$ from the edge in $G''$ (see \Cref{subsec:changpathupdateP} for how to get the path in $G'$ from each type of edge). For convenience, for each $v^{out}$, we use $S_{v^{out}}$ to denote the cluster that contains $v$. For convenience, if $v\in V(P)$, then we also define $S_{v^{out}}=\{v\}$ and $Centers[S_{v^{out}}]=\{v\}, Rep[S_{v^{out}}]=v,LU[S_{v^{out}}]=1$. 
		\begin{description}
			
			\item[Type 1: Edges inside clusters.] Each active hub broadcasts its cluster's id (each cluster has a unique cluster id, it can be the largest vertex id in the cluster, for example) to the whole graph. All the hubs in the same cluster form a clique in $G''$. 
			
			\item[Type 2: Upwards edges for upper clusters. ] For each active hub $h$, it broadcasts tokens to build a BFS tree with depth $\Theta(d\log n)$ on $G'$, denoted as $T_h$. For each $p\in P$, we define an order $\preceq'_p$ over all active hubs $h'\in T_h$ with $Centers[S_{h'}]\cap V(p)\not=\emptyset$ as follows: Recall that $Centers[S_{h'}]\cap V(p)$ can contain at most $1$ element, denote the only element as $v_{h'}$. Define $LU_p[S_{h'}]=LU[S_{h'}]$ if $v_{h'}=Rep[S_{h'}]$, and $LU_p[S_{h'}]=-1$ if $v_{h'}\not=Rep[S_{h'}]$. Then $h_1\preceq'_ph_2$ iff. $v_{h_1}\prec_pv_{h_2}$ or $v_{h_1}=v_{h_2},LU_p[S_{h_1}]\le LU_p[S_{h_2}]$. If there exists $h'\in T_h$ with $Centers[S_{h'}]\cap V(p)\not=\emptyset$, then $h$ broadcasts the partial virtual graph edge $(h,h^*)$ to the whole graph where $h^*$ is an arbitrary maximal active hub with respect to $\preceq'_p$, i.e., for any active hubs $h'\in T_h$ such that $Centers[S_{h'}]\cap V(p)\not=\emptyset$, we have $h'\preceq'_ph^*$. This can be found by aggregation on $T_h$. 
			
			\item[Type 3: Upwards edges for lower clusters. ] For each active hub $h$, it builds a reversed BFS tree $T'_h$ with depth $O(d\log n)$, i.e., all vertices in $T'_h$ can reach $h$ by a path with length at most $O(d\log n)$. For each $p\in P$, if there exists active hub $h'\in T'_h$ with $Centers[S_{h'}]\cap V(p)\not=\emptyset$, then $h$ broadcasts the partial virtual graph edge $(h^*,h)$ to the whole graph where $h^*$ is an arbitrary minimal active hub with respect to $\preceq'_p$, i.e., for any active hub $h'\in T'_h$ such that $Centers[S_{h'}]\cap V(p)\not=\emptyset$, we have $h^*\preceq'_ph'$. This can be found by aggregation on $T'_h$. 
			
			\item[Type 4: Downwards edges.] For each active hub $h$, for any $c\in Centers[S_h]$, it broadcasts $c$'s path id and position on path $PI[c],I[c]$, and two indicator $To_c[h],From_c[h]\in\{0,1,2\}$ defined by: If $c^{out}\in T_h$ then $To_c[h]=2$, otherwise if $c^{in}\in T_h$ then $To_c[h]=1$, otherwise $To_c[h]=0$; similarly if $c^{in}\in T'_h$ then $From_c[h]=2$, otherwise if $c^{out}$ in $T'_h$ then $From_c[h]=1$, otherwise $From_c[h]=0$. Now type 4 edges contains all the edges $(h_1,h_2)$ such that there exists $c_1\in Centers[S_{h_1}],c_2\in Centers[S_{h_2}]$ satisfying $PI[c_1]=PI[c_2],To_c[h_1]>0,From_c[h_2]>0$, and either $0<I[c_1]-I[c_2]<d\log n$, or $To_c[h]+From_c[h]>2,I[c_1]=I[c_2]$. Notice that in that case, $h_1$ can reach $h_2$ through a path with length bounded by $\tOh(d)$. 
			
			\item[Type 5: Terminal edges.] For each active hub $h$, if $T_h$ contains a vertex $v$ in $\{t\}\cup H\backslash H_a$, then it broadcast the edge $(h,v)$ to the whole graph. 
			
		\end{description}
		
		One can see that each vertex broadcasts at most $\tOh(\conn)$ messages to the whole graph, and the tree $T_h,T'_h$ aggregate $\conn$ messages. Since there are at most $\tOh(\conn^2\alpha p)$ active hubs, this cause a dilation $\tOh(d+D)$ and congestion $\tOh(\conn^3\alpha p)$. Now every vertex knows the same graph $G''$ locally. The following lemma shows that $G''$ is a partial virtual graph. The proof is deferred to Section~\ref{sec:clusteringAndSemi}.
		
		\begin{lemma}[Partial virtual graph; Proof in Section~\ref{subsec:partialvirtual graph}]\label{lem:HighLevelGraph}
			With high probability, the $G''$ defined by the above 5 types of edges is a partial virtual graph on $(G',T=\{t^{out}\}\cup H\backslash H_a)$ with dilation $\Theta(d\log n)$. 
		\end{lemma}

		\subsection{Step \ref{step3}: Find augmenting path.}\label{subsec:findaugmentingpath}
		
		One can see that there are four cases in Step \ref{step3}. In the first three cases, a path $p''$ should be found, in the last case, a cut should be output. We discuss each case separately in the following. Recall that $H_r$ is the set of all the active hubs that $s$ can reach in $G''$. 
		\begin{description}
			\item[Case 1 ($|H_r|=\Omega(\conn\alpha p\log n)$):] The reason that we do not uniformly sample a hub among all hubs in $H_r$ is that we might finally get a path in $G'$ with length $\omega(\conn\alpha\log n)$, which we want to avoid. Thus, instead of sampling from $H_r$, we find a subset $H'_r\subseteq H_r,|H'_r|=\Theta(\conn\alpha p\log n)$ such that $s$ can reach all the vertices in $H'_r$ through vertices in $H'_r$; and if one hub in a cluster is in $H'_r$, then all hubs in the same cluster are all in $H'_r$. This can be done since each cluster contains at most $O(\conn\alpha p\log n)$ hubs in $H_r$. To find $H'_r$, we start from $H'_r=\{s\}$, repeatedly adding $h\in H_r$ that $H'_r$ can reach, and if $h$ is added, all hubs in the same cluster as $h$ in $H_r$ are also added, until $|H'_r|=\Theta(\conn\alpha p\log n)$. Then we sample a hub $h_1\in H'_r$ uniformly at random and find $p''$ as the path using vertices in $H'_r$ from $s$ to $h_1$. All the above procedures can be done locally in each vertex since $G''$ is shared by all the vertices. We also need to guarantee that each vertex gets the same path in $G''$, this can be done by raising a leader, sampling a vertex in $H'_r$, and each vertex finding the unique path with the smallest lexicographical order to the sampled vertex. We will show how to turn this path into a path in $G'$ in \Cref{subsec:changpathupdateP}.
			
			\item[Case 2 ($t^{out}\in H_r$):] The path is $p''=s\stackrel{G''}{\to}t^{out}$. Similarly, a unique path from $s$ to $h_2$ in $G''$ is shared by all vertex. We will show how to turn this path into a path in $G'$ in \Cref{subsec:changpathupdateP}. 
			
			\item[Case 3 (there exist $h_0\in H_r$ with $h_0\in H\backslash H_a$):] We need to sample a uniformly random hub $h^*$ among all hubs $h$ with $h\preceq_\mathcal{C}h_0$. One can see that since $h_0$ is not an active hub, the number of hubs $h$ with $h\preceq_\mathcal{C}h_0$ must be $\Omega(\conn\alpha\log n)$. To sample $h$, recall that in Definition~\ref{def:ordering}, $\preceq_\mathcal{C}$ only needs the information $Rep[h_0],LU[h_0]$. Thus, $h_0$ broadcast the path id and position on the path of $Rep[h_0]$ and $LU[h_0]$. By using this information, each vertex $h$ with $h\preceq_\mathcal{C}h_0$ can mark itself. Denote the set of all the marked vertex as $L_{h_0}$. To sample one vertex, each vertex in $L_{h_0}$ samples itself with probability $1/|L_{h_0}|$, and aggregates through the whole graph to see whether there is exactly one vertex sample itself. If not, then repeat. The procedure will end with a high probability in $O(\log n)$ rounds. That shows how to sample the $h^*$ uniformly at random. The path is $p''=s\stackrel{G''}{\to}h_0\stackrel{G'}{\to}h^*$. We will see how to turn the path $s\stackrel{G''}{\to}h_0$ into a path in $G'$ in \Cref{subsec:changpathupdateP}. According to Remark~\ref{rem:orderingReachability}, there is a path from $h_0$ to $h^*$ in $G'$. However, finding the path from $h_0$ to $h^*$ is a complicated procedure, we defer it to \Cref{subsec:substepfindpathfromh0toh}. 
			
			\item[Case 4 ($H_r$ do not contain any $t^{out}$ or a non-active hub):] We will need the following claim. 
			
			\begin{claim}\label{cla:reachall}
				$\cup_{h\in H_r}T_h$ contains all the vertices that $s^{out}$ can reach in $G'$ with high probability. 
			\end{claim}  
		
			To see this, suppose $v\in V'$ is reachable from $s$ but not in $\cup_{h\in H_r}T_h$, then according to Lemma~\ref{lem:HighLevelGraph} and Definition~\ref{def:semivirtual graph}, $s^{out}$ should reach a vertex in $T=\{t^{out}\}\cup H\backslash H_a$, which is a contradiction. Thus, according to Lemma~\ref{lem:correspondingCut}, by finding all the vertex $v\in V$ that $v^{out}$ is in $\cup_{h\in H_r}T_h$, we get a vertex cut with size at most $\conn$. 
			
		\end{description}

		\subsection{Substep: Find path from $h_0$ to $h^*$.}\label{subsec:substepfindpathfromh0toh}
		We first describe a lemma showing how to find a path between two vertices in the same cluster $S\in\mathcal{S}$.
		
		\begin{lemma}[Find path in cluster; Proof in Appendix~\ref{app:mergeClusters}]\label{lem:FindPath}
			On an undirected graph $H=(V,E)$ with diameter $D$, given an connected induced subgraph $H[V']$ of $H$, and two vertices $s,t\in V'$, given an integer $1\le d\le N$, there exists a CONGEST model algorithm on $H$ finding a path from $s$ to $t$ in $H[V']$, with dilation $\tilde{O}(d+D)$ and congestion $\tilde{O}(|V|/d)$. 
		\end{lemma}
		
		For two vertices $u,v\in S$, we use $Path_{S}(u,v)$ to denote the path found by Lemma~\ref{lem:FindPath} from $u$ to $v$ in $S$, using $H_S$ as the communication graph $H$ in the lemma. For $p\in P$, Let the path $p'$ in $G'$ be the corresponding backward path of $p$. We use $Path_{p'}(u,v)$ for two vertices $u,v$ in $p'$ to denote the subpath from $u$ to $v$ in $p'$. Such a path can be found distributively by broadcasting the position of $u,v$, all vertices in the path with position between $u,v$ make an edge towards $v$. In the following, we consider three cases and show how to find the path from $h_0$ to $h^*$ in each case. For convenience, we assume $M(h_0),M(h^*)\not\in V(P)$, and the cluster containing $M(h_0),M(h^*)$ are $A,C$ separately. The case when $M(h_0)$ or $M(h^*)$ is in $V(P)$ only makes things easier which can be treated similarly. In the following, we will also define cluster $B$. For convenience, we assume the size of $A,B,C$ are all bounded by $O(\conn\alpha\log n)$, which guarantees the length of the path. We will discuss what should we do when one of $A,B,C$ has size $\Omega(\conn\alpha\log n)$ in step 4. Let $a=Rep[A],c=Rep[C]$ and let $a'\in A,c'\in C$ be the vertices such that $a'$ is a neighbor of $a$, $c'$ is a neighbor of $c$. Let $p$ be the path that $a,c$ is in.
		
		\begin{description}
			\item[Case 1 ($A$ is an upper cluster):] Suppose $A$ is connected to cluster $B$ through edge $(a_0,b_0)$, with $b\in Centers[B]$ and $a\prec_pb$. Let $b'$ be the vertices in $B$ that are neighbors of $b$. Then the path is $Path_A(h_0,a_0^{out})\to b_0^{out}\to Path_B(b_0^{out},b'^{out})\to b^{in}\to  Path_{p'}(b^{in},c^{out})\to c'^{out}\to Path_C(c'^{out},h^*)$. Note that according to the definition of upper cluster, it might be the case that $A$ is connected to a vertex $b\succ_pa$ on the path (but not a cluster). In that case, the problem becomes easier: there is an edge directly from $a_0^{out}$ to $b^{in}$. 
			
			\item[Case 2 ($A$ is a lower cluster and $c\prec_pa$):] $A$ can go to its representative and then following the backwards of the path to reach $C$. The path is $Path_A(h_0,a'^{out})\to a^{in}\to Path_{p'}(a^{in},c^{out})\to c'^{out}\to Path_C(c'^{out},h^*)$.
			\item[Case 3 ($A$ is a lower cluster and $a=c$):] In that case, $C$ must be a lower cluster. Suppose $C$ is connected to cluster $B$ through edge $(b_0,c_0)$, with $b\in Centers[B]$ and $b\prec_pc$. Let $b'$ be the vertices in $B$ that is a neighbor of $b$. Then the path is $Path_A(h_0,a'^{out})\to a^{in}\to Path_{p'}(a^{in},b^{out})\to b'^{out}\to Path_B(b'^{out},b_0^{out})\to c_0^{out}\to Path_C(c_0^{out},h^*)$. Note that according to the definition of lower cluster, it might be the case that $C$ is connected to a vertex $b\prec_pc$ on the path (but not a cluster). In that case the problem become easier: there is an edge from $b^{out}$ to $c_0^{out}$. 
		\end{description}
		
		On the case that one of $A,B,C$ has size $\Omega(\conn\alpha)$, we need the following lemma. The proof the lemma is deferred to Appendix~\ref{app:mergeClusters}.
		
		\begin{lemma}[Find small piece in cluster; proof in Appendix~\ref{app:mergeClusters}]\label{lem:FindCluster}
			On an undirected graph $H=(V,E)$ with diameter $D$, given an induced subgraph $H[V']$ of $H$ satisfying $|V\backslash V'|\le k$, $H[V']$ is connected. Given a vertex $s\in V'$, given an integer $x$, there exists a CONGEST model algorithm on $H$ finding a vertex set $V''\subseteq V'$ such that $s\in V''$, $|V''|=\Theta(x)$, $H[V'']$ is connected, with dilation $\tilde{O}(kD)$ and congestion $\tilde{O}(k)$. 
		\end{lemma}
		
		Without loss of generality, suppose the first of $A,B,C$ that has size $\Omega(\conn\alpha)$ is $A$. Then when we want to find the path in $A$ by Lemma~\ref{lem:FindPath}, we first use Lemma~\ref{lem:FindCluster} to find a connected induced subgraph $A'$ inside $A$ with $\Theta(\conn\alpha)$ vertices containing $h_0$, then sample a vertex $t^*$ in $A'$ uniformly at random, then use Lemma~\ref{lem:FindPath} to find the path from $h_0$ to $t^*$ in $A'$. Instead of reaching $h^*$, we reach $t^*$ as the final vertex. Same if the first of them is $B$ or $C$: we stop at the point when we want to find a path inside $B$ or $C$, and end at $t^*$ inside $B$ or $C$. 
		
		Since each cluster has size bounded by $\tOh(\conn\alpha)$, the above procedure has dilation $\tOh(d+D)$, congestion $\tOh(\conn\alpha/d)$. The length of the path from $h_0$ to $h^*$ is bounded by $\tOh(\conn\alpha)$.
		
		\subsection{Step \ref{step4}: Change path in $G''$ to $G'$ and update $P$.}\label{subsec:changpathupdateP}
		
		Now given a path from $s$ to $h'$ in $G''$ ($h'$ can be $h_1,t^{out}$ or $h_0$ in step 3), we want to turn it into a path in $G'$. We first refine this path such that it goes into each cluster (hubs with the same cluster IDs, which form a clique) and go out at most once: we put an edge from the first vertex on the path intersecting the clique to the last vertex on the path intersecting the clique and discard the subpath between them. Now we replace each edge $(h_L,h_R)$ in $p''$ into a path in $G'$. We consider three cases according to the type of $(h_L,h_R)$.
		\begin{description}
			\item[Case 1 (Type 1 edge):] It this case, $h_L,h_R$ are in the same cluster, denoted as $A$. The path becomes $Path_A(h_L,h_R)$ in $G'$, which can be found with dilation $\tOh(d+D)$ and congestion $\tOh(|A|/d)$ using Lemma~\ref{lem:FindPath}. Let $S[A]$ be the number of hubs inside $A$, since we sample each hub with probability $p$, w.h.p we have $|A|=\tilde{O}(S[A]/p)=\tilde{O}(S[A]d)$. Notice that $Path_A(h_L,h_R)$ exists at most once for each $A$, and there are $\tOh(\conn\alpha p)$ active hubs inside $H_r$ (or $H'_r$), which means all type 1 edges can be turned into paths with dilation $\tOh(d+D)$ and congestion $\tOh(\conn\alpha p)$, while the total length of these paths is bounded by $\sum|A|=\tilde{O}(\sum S[A]d)=\tilde{O}(\conn\alpha)$.
			
			\item[Case 2 (Type 2,3,4,5 edge):] Notice that in this case, $h_L$ has distance bounded by $\tOh(d)$ to $h_R$. The path from $h_L$ to $h_R$ in $G'$ can be found by growing a BFS tree with depth at most $\tOh(d)$. Finding all of them cause dilation $\tOh(d)$ and congestion $\tOh(\conn\alpha p)$, and the sum of the length of all these paths is bounded by $\tOh(\conn\alpha)$.
		\end{description} 
		
		The total dilation is $\tOh(d+D)$, the total congestion is $\tOh(\conn\alpha p)$, and the total length of the first part path is $\tOh(\conn\alpha)$.  
		
		Now we get a path in $G'$ but not a simple path. We use Lemma~\ref{lem:pushlabel} on the path to find the position of each vertex on the path. If a vertex is repeated, it only preserves the edge point out to the vertices with the largest position id (farthest from $s$), and deletes other out edges. Now all the edges form a subgraph, where the connected components of the subgraph containing $s$ is a simple path. The following lemma shows how to find it. The proof is deferred to Appendix~\ref{app:mergeClusters}.
		
		\begin{lemma}[Turn path into simple path; Proof in Appendix~\ref{app:mergeClusters}]\label{lem:SimplePath}
			On the network $G=(V,E)$ with undirected diameter $D$, there exists a CONGEST model algorithm given the edge direction, a vertex $s$, a subgraph $H$ with $\mu$ vertices which contains a path $p$ starting from $s$, satisfying there are no edges in $H$ between vertices in $p$ and other vertices in $H$, an integer $1\le d\le\mu$, output for each vertex whether it is in $p$ or not; the algorithm has dilation $\tilde{O}(d+D)$ and congestion $\tilde{O}(\mu/d)$.
		\end{lemma}
		
		We use Lemma~\ref{lem:SimplePath} to fix the final simple path. Now we get a simple path $p'$ on $G'$. Then we check whether the internal points intersect with the endpoints of any path in $P$, if there exists, then we discard all the vertices after the intersection. According to Lemma~\ref{lem:augmentingConnectivity}, by computing $\oplus(E(M(p'))\cup E(P))$ locally, and use Lemma~\ref{lem:SimplePath}, we get $k+1$ internally vertex disjoint paths staring from $s$. Since $p$ has length bounded by $\tOh(\conn\alpha)$, the step has dilation $\tOh(d+D)$ and congestion $\tOh(\conn\alpha/d)$. 
		\subsection{Analysis of Algorithm}\label{subsec:largealphaanalysis}
		\paragraph{Round complexity.} The algorithm is described above. As shown in each step of the algorithm details, the total dilation and congestion for all $\conn$ loops are bounded by $\tOh(\conn d+\conn^3D)$ and $\tOh(\conn^4\alpha/d)$. Recall that $d=\kappa^{1.5}\sqrt{n}$, which leads to the dilation $\tOh(\conn^{2.5}\sqrt{n}+\conn^3D)$ and congestion $\tOh(\conn^{2.5}\alpha/\sqrt{n})$ claimed in~\Cref{lem:largeAlphaAlgorithm}.
		
		\paragraph{Correctness.} Similarly to  \Cref{subsec:proofofloweralpha}, We first prove that, if a cut is output, it must be a valid cut. Recall in the algorithm, a cut is output iff. it falls into the last sentence of Step~\ref{step3}. Since $S'$ contains no outgoing edges, $S'$ contains all the vertices that $u^{out}$ can reach. According to~\Cref{lem:correspondingCut}, $\{v\in V\mid v^{in}\in S',v^{out}\not\in S'\}$ is a cut as long as $V'\backslash S'\not=\emptyset$. The latter claim is because $t\not\in S'$.
		
		Now suppose there exists $L\subseteq V$ such that $|\partial(L)|<\conn,\{s,t\}\cap \partial^+_v(L)=\{s\},|L|\le\alpha$, we will prove that $\bot$ will be output with at most constant probability. According to ~\Cref{cla:reachall}, $\bot$ will not be output at Step~\ref{step3} with high probability, so we only consider the case that $\bot$ is output at the last line of the algorithm. Let $R=V\backslash\partial^+(L)$. If all paths in $P$ end at $R$, then according to Lemma~\ref{lem:augmentingConnectivity}, there exists $\conn$ internally vertex disjoint path between $s$ and $R$, which contradiction the fact that $|\partial(L)|<\conn$. 
		
		Let $I_i$ be the event that the first loop where the ending set of $P$ contains a vertex not in $R$ is loop $i$. We will bound the probability that $I_i$ happens. Let $P$ be the flow-path before the $i$-th loop. According to the algorithm description, at the $i$-th loop, it finds an augmenting path $p'$ in $G'$ either ending at an endpoint of $P$ (we denote all the endpoints of $P$ as $T$), or ends at $t^{out}$ or $h_1,h^*$ or $t^*$. Denote $R=V\backslash \partial^+_v(L)$. Note that $I_i$ implies $T\subseteq R$. $M(p')$ ends at a point in $R$ with probability at least $1-\frac{1}{2\conn}$. We consider each case.
		\begin{enumerate}
			\item If $p'$ ends at $T$ or $t^{out}$, then we are done since $T\subseteq R$ and $t\in R$. 
			\item If $p'$ ends at $h^*,h_1$. Recall that $h^*,h_1$ is an uniformly random hub from $\Omega(\conn\alpha p\log n)$ hubs. Since we sample each vertex as hub with probability at least $p$ inside $L$, with high probability, the hubs inside $\partial^+_v(L)$ is bounded by $O((\alpha+\conn)p\log n)$ with high probability. Since $\conn<\alpha$, we have $M(h_1),M(h^*)\in R$ with probability at least $1-\frac{1}{2\conn}$.
			\item If $p'$ ends at $t^*$. Recall that $t^*$ is a uniformly random vertex from $\Omega(\conn\alpha)$ vertices. Thus, $M(t^*)\in R$ with probability at least $1-\frac{1}{2\conn}$.
		\end{enumerate}
		
		The event $I_i$ happens with probability at most $\frac{1}{2\conn}$ according to the above discussion, which means $\bot$ is output with probability bounded by a constant. 
		
		\section*{Acknowledgment}
		
		We would like to thank Danupon Nanongkai for numerous fruitful discussions through out the project, and the reviewers for their meticulous reading and comments.
		\bibliography{./ref}
		
		\appendix
		
		\section{Clustering and Partial Virtual Graph}\label{sec:clusteringAndSemi}
		In this section we prove two main gradients of our algorithm for large $\alpha$: the clustering algorithm in Lemma~\ref{lem:clustering} and the partial virtual graph properties in Lemma~\ref{lem:HighLevelGraph}.
		
		\subsection{Proof of \Cref{lem:clustering}}\label{subsec:clustering}
		
		We give the algorithm for building a paths centered clustering on $(G,P)$. Each vertex in $V(P)$ broadcast tokens to build a BFS tree: initially only vertices in $V(P)$ are activate, on each round all active vertices send a token to all its neighbors if it haven not done so. At the end of each round, if a vertex that have not been activated receive a token (probability more than one token), then it joins the BFS tree of an arbitrary vertex that sent it the token, then become activate. The procedure has dilation $O(D)$ and congestion $O(1)$, since each active vertices only sends once, and each vertex has distance at most $D$ to a vertex in $V(P)$. 
		
		Now we get a partition of $V$, each part is a tree rooted at a vertex $v\in V(P)$ with depth at most $D$, we denote it as $T_v$. Each subtree rooted at a child of $v$ on $T_v$ is a cluster $S$. All $S$ form a partition of $V\backslash V(P)$, and $G[S]$ has diameter $D$. Now we want to combine clusters in order to make each cluster either an upper cluster or a lower cluster. We maintain the center set $Centers[S]$ for each cluster $S$. Initially $Centers[S]=\{v\}$, where $T_v$ is the tree containing $S$. We write $H_S=(S\cup Centers[S],\{(u,v)\mid u\in S,v\in S\cup Centers[S]\})$. One can see that $H_S$ has diameter $D$ initially. During the algorithm we will maintain the property that $H_S$ has diameter $kD$. For each vertex $u\in V\backslash V(P)$, we use $S_u$ to denote the cluster $u$ is in. 
		
		For path $p_i=(v_0,v_1,...,v_j,...,v_{\ell})$, we define $PI[v_j]=i,I[v_j]=j$ to denote the path id and position on path for $v_j$. For each edge $(u,v)$ where $u,v\not\in V(P),S_u\not=S_v$, if there do not exists $c_1\in Centers[S_u],c_2\in Centers[S_v]$ such that $PI[c_1]=PI[c_2],I[c_1]\not=I[c_2]$, then we call edge $(u,v)$ a \emph{critical edge}. The idea of the algorithm is to eliminate all critical edges. 
		
		We run several phases until there are no critical edge. At each phase, each cluster tosses a coin, getting head or tail with equal probability $\frac{1}{2}$. Clusters use the shortcuts $H_S$ to share information inside a cluster. For a critical edge $(u,v)$ where $S_u$ has tail and $S_v$ has head, $S_u$ want to join $S_v$. Let $S'$ contains all the clusters that want to join $S_v$. We run at most $k$ times the following loops. For a cluster $S$, define $PI[S]=\{PI[c]\mid c\in Centers[S]\}$. At each loop, $S_v$ try to find $i\in (\cup_{S\in S'}PI[S])\backslash PI[S_v]$, i.e., path id in $S'$ but not in $S_v$. This can be done by making  congestion $k$. If $i$ do not exists, then loops are ending. Otherwise, pick arbitrary $S\in S'$ with $i\in PI[S]$, merge $S$ to $S_v$. After merging, the new center of $S_v$ becomes $Centers[S_v]\cup Centers[S]$. One can see that $Centers[S_v]\cup Centers[S]$ still contains at most $k$ elements, according to the definition of critical edge (there do not exist two different elements with the same path id). Also notice that each vertex in $S_v$ has distance at most $D$ to a vertex in $Centers[S_v]$, thus, $H_{S_v}$ has diameter $kD$. After merging, some critical edges might no longer be a critical edges. Recompute all critical edges and $S'$ (note that we do not re-toss the coin), and continue the next loop. Since each loop increase the size of $Centers[S_v]$ by $1$, there can be at most $k$ loops. After all the loops, all clusters $S\in S'$ satisfying $PI[S]\subseteq PI[S_v]$. Now all clusters in $S'$ merge to $S_v$, after which the phase ends. 
		
		At each phase, a critical edge with tail on one side and head on another side must disappear at the end of this phase. Moreover, if an edge is not a critical edge, then it will never become a critical edge. That is because an edge is not a critical edge if either one of its end points is in $V(P)$, or the centers of both side shares the same path id with different position on the path, and $Centers[S]$ never delete elements. Each critical edge disappear with constant probability at each phase, thus, after $O(\log n)$ phases, with high probability, there are no critical edges. Each phase contains at most $k$ loops, each loop contains a broadcasting inside shortcut with $k$ messages and diameter $kD$. Thus, the algorithm has dilation $\tilde{O}(k^2 D)$ and congestion $\tilde{O}(k^2)$. 
		
		At the end, $Centers[S]$ contains at most one vertex in each path, and $H_S$ form a shortcut for $S$ with diameter $kD$. Moreover, all the edges adjacent $(u,v)$ with $u\in S,v\not\in S$ is not a critical edge. Consider several cases:
		\begin{enumerate}
			\item If there exists $(u,v)$ such that $v\not\in V(P)$, then there exists $c_1\in Centers[S_u],c_2\in Centers[S_v]$ such that $PI[c_1]=PI[c_2],I[c_1]\not=I[c_2]$. We let $c_1$ be the representative of $S$, denoted as $Rep[S]$, and set $S$ as upper cluster or lower cluster according to whether $I[c_1]<I[c_2]$ or $I[c_2]<I[c_1]$.
			\item Otherwise, if there exists $(u,v_1),(u,v_2)$ such that $v_1,v_2\in V(P),PI[v_1]=PI[v_2],I[v_1]\not=I[v_2]$. We assume $v_1\in Centers[S]$, otherwise consider two cases: if $Centers[S]$ contains a center $c$ with $PI[c]=PI[v_1]$, then let $v_1=c$ and $v_2$ be another vertex among $v_1,v_2$; if $Centers[S]$ do not contains such a center, then we add $v_1$ to $Centers[S]$. In any case, we can assume $v_1\in Centers[S]$. Let $v_1$ be the representative of $S$, and set $S$ as upper cluster or lower cluster according to whether $I[v_1]<I[v_2]$ or $I[v_2]<I[v_1]$.
			\item Otherwise, all the edges adjacent to $S$ is like $(u,v)$ where $v\in V(P)$ and $v$ only appear once for each path. In that case, there are at most $k$ vertices in $\partial(S_i)$. We return $\partial(S_i)$ as a vertex cut.
		\end{enumerate}
		
		\subsection{Proof of \Cref{lem:HighLevelGraph}}\label{subsec:partialvirtual graph}
		
		We will prove several claims that shows properties about $G''$, which will help us prove the Lemma. Recall that we define $S_{v^{out}}$ as the cluster in $\mathcal{S}$ containing $v$. For convenience, if $v\in V(P)$, then we also define $S_{v^{out}}=\{v\}$ and $Centers[S_{v^{out}}]=\{v\}, Rep[S_{v^{out}}]=\{v\},LU[S_{v^{out}}]=1$, and $S_{v^{out}}$ is treated as an upper cluster (since it can reach $v^{out}$). We define $h_1\prec'_ph_2$ for two hubs $h_1,h_2$ as the following: there exists $c_1\in Centers[S_{h_1}]\cap V(p),c_2\in Centers[S_{h_2}]\cap V(p)$ and $c_1\prec_p c_2$. Recall that $T=\{t\}\cup H\backslash H_a$, and our goal is assuming a path from $s$ to a vertex $v\in V'$ exists in $G'$, proving that there is a path from $s$ to $T$ in $G''$ or $s$ to $v'$ in $G''$ where $v'$ has distance $\Theta(d\log n)$ to $v$..
		
		\begin{claim}\label{cla:havehubs}
			with high probability, for each cluster $S$ that contains at least one hub, there exists a hub in the cluster that has distance at most $\tOh(d\log n)$ to any vertex $c\in Centers[S]\cup S$. Moreover, any consecutive sub path of $P$ with length $\Omega(d\log n)$ contains at least one hub.
		\end{claim}  
		\begin{proof}
			The first sentence is because $G[S]$ is connected and also connect to $c$. In the induced subgraph $G[S\cup \{c\}]$, we find all the vertices with distance $\Theta(d\log n)$ to $c$. There are at least $\Omega(d\log n)$ such vertices (or it covers the whole $S_i$, in which case it must contain the hub in the cluster), and with high probability one vertex is sampled as hub because we sample a hub with probability $p=1/d$. This same argument holds for a consecutive sub path $P$ with length $\Omega(d\log n)$. Since there are at most $n$ clusters and $n^2$ sub paths, by using union bound we get the desired conclusion. 
		\end{proof}
		\begin{claim}\label{cla:reachlowerhub}
			For any two hubs $h_1\in H_a,h_2\in H$, if there exists $p\in P$ such that $h_2\prec'_ph_1$, then $h_1$ can reach either $T$ or $h_2$ in $G''$. 
		\end{claim}
		\begin{proof}
			For any active hub $h$ satisfying $Centers[S_{h}]\cap V(p)\not=\emptyset$, denote the only element in $Centers[S_{h'}]\cap V(p)$ as $c_{h}$. Let the set $H^*$ contains all the active hubs $h$ that $h_1$ can reach in $G''$ such that $c_{h_2}\prec_pc_h$, let $h^*$ be one of the minimal in $H^*$ with respect to $\preceq'_p$, i.e., $h^*\preceq'_p h$ for any $h\in H^*$. Firstly, $H^*$ is not an empty set, since $h_1$ is in $H^*$. We claim that $I[c_{h^*}]-I[c(h_2)]=O(d\log n)$. Otherwise, for convenience we suppose $M(h^*)\in V\backslash V(P)$ and is in cluster $S$. The case $M(h^*)\in V\backslash V(P)$ only make things easier. Then with high probability there exists an active hub $c\in S$ which has distance at most $O(d\log n)$ to $c_{h^*}$ according to \Cref{cla:havehubs}, and $h^*$ has an edge to $c$ according to type 1 edge. There must exist a active hub $h_3$ in $V(P)$ such that $I[c_{h^*}]>I[M(h_3)]>I[c_{h_2}]$ and $I[c_{h^*}]-I[M(h_3)]=O(d\log n)$, according to \Cref{cla:havehubs}. According to the definition of type 4 edge, $c$ has an edge to $h_3$, which means $h_3$ can be reached from $h_1$, leading to a contradiction. Thus, we have $I[c_{h^*}]-I[c_{h_2}]=O(d\log n)$. For convenience we assume $M(h_2)$ is in a cluster $S'$, the case that $M(h_2)$ is in $V(P)$ only make things easier. Let $c'$ be the hub in $S'$ that has distance $O(d\log n)$ to $c_{h_2}$. If $c'\in T$, then $h^*$ can reach $c'$ by first use type 1 edge to reach $c$, then use type 5 edge to reach $c'$ since $c$ has distance at most $O(d\log n)$ to $c'$. Otherwise $c'$ is an active hub, and we can use type 4 edge to reach $c'$, then use type 1 edge to reach $h_2$. 
		\end{proof}
		\begin{claim}\label{cla:reduceedge}
			For $h_1,h_2\in H_a$, if $h_2\in T_{h_1}$, then either $h_1$ can reach $T$, or $h_1$ can reach $h_2$ in $G''$, with high probability. 
		\end{claim}
		\begin{proof}
			Let $p\in P$ be the path that $Rep[S_{h_2}]$ is in. For any $h\in H_a$, we define $v_{h}$ as the only vertex in $Center[S_h]\cap V(p)$. If $T_{h_1}$ contains a vertex in $T$ then we are done. Otherwise, according to the definition of type 2 edge, there exists an edge $(h_1,h^*)$ such that $v_{h_2}\prec_p v_{h^*}$ or $v_{h_2}=v_{h^*},LU[S_{h_2}]\le LU[S_{h^*}]$ and $h^*$ is an active hub. In the case that $v_{h_2}\prec_p v_{h^*}$, we are done according to \Cref{cla:reachlowerhub}, since $h_2\prec'_p h^*$. So we only need to consider the case $v_{h_2}=v_{h^*},LU[S_{h_2}]\le LU[S_{h^*}]$, we denote $v_{h_2},v_{h^*}$ as $c$. Recall that $LU[S_{h_2}]>0$ and is $1$ or $2$ if $S_{h_2}$ is a lower or upper cluster. Thus, we have $LU[S_{h^*}]>0$, which means $c=Rep[S_{h^*}]$. Now consider two cases.
			\begin{enumerate}
				\item If $LU[S_{h^*}]=2$, in which case $S_{h^*}$ is an upper cluster. Then according to the definition of upper cluster, there exists an edge $(v_1,v_2)$ where $v_1\in S_{h^*}$ and $v_2$ is in cluster $B$ with $b\in Center[B]$ and $c\prec_p b$. According to \Cref{cla:havehubs}, there is an active hub $h'\in S_{h^*}$ that has distance at most $O(d\log n)$ to $v_1$. Consider the path $p_0$ starting from $v_2$, go through the connected induced subgraph $G'[B]$, to $b^{in}$ and go back to $c^{out}$ through the reversed path of $p$. If $p_0$ has length bounded by $O(d\log n)$, then we have $c^{out}\in T_{h'}$, which means $h'$ is connected to a vertex in the cluster $S_{h_2}$ according to type 4 edge; otherwise, the length of $p_0$ is $\Omega(d\log n)$, in which case there is a hub $h_{p_0}$ which is an internal vertex of $p$, such that $h_{p_0}\in T_{h'}$ w.h.p. Notice that either $h_{p_0}\in B$ or $h_{p_0}\in p$, in both cases we have $h_2\prec'_ph_{p_0}$. Thus, there is a type 2 edge $(h',h^*_{p_0})$ such that $h_2\prec'_ph^*_{p_0}$ according to the definition of type 2 edge. According to \Cref{cla:reachlowerhub}, we are done.
				\item If $LU[S_{h^*}]=1$, then both $S_{h^*},S_{h_2}$ are lower cluster. According to the definition of lower cluster, there exists an edge $(v_1,v_2)$ where $v_1\in S_{h_2}$, $v_2$ is in a cluster $B$ with $b\in Center[B]$ and $b\prec_pc$. According to \Cref{cla:havehubs}, there is an active hub $h'\in S_{h_2}$ that has distance at most $O(d\log n)$ to $v_1$. Consider the path $p_0$ starting from $c^{in}$, go back to $b^{out}$ though the reversed path of $p$, and go to $v_2$ in the connected induced subgraph $G'[B]$. If $p_0$ has length bounded by $O(d\log n)$, then we have $c^{in}\in T_{h_2}$, which means there is a vertex in $S_{h^*}$ that connects to $h'$ according to type 4 edge; otherwise, the length of $p_0$ is $\Omega(d\log n)$, in which case there is a hub $h_{p_0}$ which is an internal vertex of $p$, such that $h'\in T_{h_{p_0}}$, w.h.p. Notice that either $h_{p_0}\in B$ or $h_{p_0}\in p$, in both cases we have $h_{p_0}\prec'_ph^*$. Thus, there is a type 3 edge $(h^*_{p_0},h')$ such that $h^*_{p_0}\prec'_ph^*$ according to the definition of type 3 edge. According to \Cref{cla:reachlowerhub}, we are done. 
			\end{enumerate}
		\end{proof}

		Now we are ready to prove the lemma. Suppose $s$ can reach $v\in V'$ and $p$ is the path from $s$ to $v$ in $G'$. With high probability, there exists a sequence of hubs on $p$ like $(s=h_0,h_1,...,h_k)$ such that $h_{i+1}\in T_{h_i}$ for any $0\le i\le k-1$, and $v\in T_{h_k}$: that is because every consecutive $\Theta(d\log n)$ vertices must contain a hub with high probability. According to \Cref{cla:reduceedge}, for each active hub $h_i$, either it can reach $h_{i+1}$ in $G''$, or it can reach $T$. Therefore, either $s$ can reach $T$, or $s$ can reach $h_k$, and $h_k$ can reach $v$ with distance at most $\Theta(d\log n)$.
		
		\section{Proof of Vertex Residual Graph Lemmas}\label{app:residualGraph}
		In this section we prove Lemma~\ref{lem:augmentingConnectivity} and Lemma~\ref{lem:correspondingCut}, which relate vertex residual graph to vertex connectivity.

		\begin{proof}[Proof of Lemma~\ref{lem:augmentingConnectivity}]
			We prove it by induction on the length of $p'$. When the length is $0$, the lemma trivially holds. Now suppose $|p'|=k$ and the lemma holds for $|p'|<k$. We consider two cases.
			\begin{enumerate}
				\item Suppose $p'$ ends at $v^{out}$ for some $v\not\in V_I(P)$, and $p'$ is like $(u,...,v',v)$. We use the induction on the path $p''=(u,...,v')$ to get vertex disjoint paths $P''$, which contains a path ending at $M(v')$. We will argue that by extending the path ending at $M(v')$ to $v$, we get $P'$ which satisfies all we want. There are three things to show: 1. $E(P')$ are internally vertex disjoint paths. Since $E(P'')$ are already internally vertex disjoint, we only need to show: $M(v')$ do not intersect with any other paths unless $M(v')=u$, and $v$ do not intersect with any other internal vertices. 2. $P'$ are simple paths. 3. $E(P')$ is indeed the connected component of subgraph $\oplus(E(P)\cup E(M(p'))$ containing $u$, and other components are circle. 
				
				First we show that $M(v')$ do not intersect with any other paths in $P''$ unless $M(v')=u$: firstly, we have $M(v')\not\in T$ since $V_I(p')$ cannot contain any vertices in $T$, thus, since $P''$ ends at multiset $T\cup\{M(v')\}$, exactly one path ends at $M(v')$, and $M(v')$ is not the internal vertices of any other paths according to induction hypothesis.  
				
				Now We consider two cases, differed by $v\in T$ or not. Suppose $v\not\in T$. We first show that $v\not\in V(P)$. We know $v\not\in V_I(P)$, and we also have $v\not\in T$, and $v\not =u$ since $p$ is a simple path with length at least $1$. $v$ is also not in $V(p'')$, since $p'$ is a simple path. Thus, $v$ is not adjacent to any edge in $\oplus(E(P)\cup E(M(p''))$. Therefore, by extending the path ending at $M(v')$ to $v$, all paths are still internally vertex disjoint, since we have shown $M(v'),v$ do not intersect other paths. Components relationship does not change.
				
				Suppose $v\in T$. Then $v$ is not the internal vertices of any path in $P''$ according to induction hypothesis. Thus, by adding $(M(v'),v)$ to the path ending at $M(v')$ in $P''$, we still get internally vertex disjoint paths. Components relationship does not change.
				
				Finally, $P'$ satisfies $E(P')=\oplus(E(P)\cup E(M(p')))$ since $E(P')=E(P'')\cup\{(M(v'),v)\}$ and $E(M(p'))=E(M(p''))\cup\{(M(v'),v)\}$. 
				
				\item Suppose $p'$ ends at $v^{out}$ for some $v\in V_I(P)$. According to the definition of residual graph, and the internal of $p'$ do not intersect $T$, there must exists $p\in P$ such that $(v,w)\in E(p),w\not\in T$, and $p'$ must be like $(u,...,w',w^{in},v^{out})$. Now consider two cases.
				
				Suppose $w'=w^{out}$. Use the induction on path $p''=(u,...,w'=w^{out})$ to get internally vertex disjoint paths $P''$, which contains a path ending at $w$. Since $p'$ is a simple path, $(w^{in},v^{out})$ is not in $E(p'')$. Thus, $(w,v)\not\in M(p'')$, because $(w^{in},v^{out})$ is the only edge that can be mapped to $(w,v)$. Since $(v,w)\in E(P)$, we have $(v,w)\in \oplus(E(P)\cup E(M(p'')))$. $(v,w)$ is also in $E(P'')$ since $w$ is in the same components as $u$. Thus, there exists a path ending at edge $(v,w)$ in $P''$. By deleting this edge $(v,w)$, we get $P'$. Since we only delete an edge, all properties about $P''$ still holds.
				
				When $w'\not=w^{out}$. We use the induction on path $p''=(u,...,w')$ to get internally vertex disjoint paths $P''$, which contains a path ending at $M(w')$. According to the same argument as 1, we know there is exactly one path in $P''$ that ends at $M(w')$. Since $p'$ is a simple path, $(w^{in},v^{out}),(w',w^{in})$ are not in $E(p'')$. Thus, $(w,v)\not\in E(M(p''))$, since $(w^{in},v^{out})$ is the only edge that can be mapped to $(w,v)$. Therefore, we have $(v,w)\in \oplus(E(M(p''))\cup E(P))$. If $(w,M(w'))\in \oplus(E(M(p''))\cup E(P))$ also holds, since $M(w')$ is an end point of $P''$, one path in $P''$ must be like $(u,...,v,w,M(w'))$. By deleting the last two edges of the path to $(u,...,v)$, we get $P'$, which maintains all the properties of $P''$. Thus, the only case left is $(w,M(w'))\not\in E(M(p'')) +  E(P)$. In that case we need to do path shifting. We consider two cases, differed by whether $(v,w)$ is in $E(P'')$ or not.
				
				If $(v,w)\in E(P'')$, suppose $p_v$ is the path in $P''$ containing $(v,w)$. Let $p_w$ be the unique path in $P''$ ending at $M(w')$. We consider two cases. Suppose $p_v=p_w$, then by adding edge $(M(w'),w)$ and delete $(v,w)$ from $E(P'')$, we get a path ending at $v$ (which is a sub path of $p_v$ by delete the last edge of $p_v$), as well as a circle, containing the sub path of $p_v$ starting from $w$ to $M(w')$ and edge $(M(w'),w)$. Since $p_v$ is a simple path, the circle is vertex disjoint, and the new path is simple. If $p_v\not=p_w$, then we shift this two paths by adding $(M(w'),w)$ and delete $(v,w)$ from $E(P'')$. One of the new path take the whole part of $p_v$ and the latter sub path of $p_w$ starting from $w$; the other new path take the former sub path of $p_w$ end at $v$. All properties hold. 
				
				If $(v,w)\not\in E(P'')$, then according to induction hypothesis, $(v,w)$ is in a vertex disjoint circle. by adding $(M(w'),w)$ and delete $(v,w)$, we add the circle into the path and get a new path ending at $v$. All properties hold.

			\end{enumerate}
		\end{proof}

		\section{CONGEST Algorithms Based on Merging Clusters}\label{app:mergeClusters}
		In this section we prove missing lemmas stated in Section~\ref{sec:largeAlpha}. They are all based on techniques that start from clusters with single vertices, merge clusters while maintaining cluster properties, and finally combined all clusters into a single cluster.

		\begin{proof}[Proof of Lemma~\ref{lem:SimplePath}]
			We maintain clusters $\mathcal{S}$, initially each cluster $S\in\mathcal{S}$ contains exactly one vertex in $H$. We run several phases, at each phase, each cluster toss a coin, getting head or tail. In each phase, tail clusters merge to head clusters if they share an edge in $H$. Each cluster also maintain the information whether it contains $s$. Communication inside each cluster happened by broadcasting in each cluster if the cluster has size at most $d$, or broadcasting through the whole network if it has size more than $d$. Since there are at most $\mu$ vertices in $H$, the number of clusters that broadcast through the whole network is bounded by $\mu/d$. Since at each phase, each edge has constant probability to merge two clusters (and become an edge inside a cluster), after $O(\log n)$ phases, there are no edge connecting two different cluster. Now, path $p$ is the cluster that contain $u$.
		\end{proof}
		
		\begin{proof}[Proof of Lemma~\ref{lem:pushlabel}]
			If a vertex is repeated $r$ times, we just treat it as $r$ vertices, in which case we can treat $p$ as a simple path. We maintain clusters $\mathcal{P}$, initially each cluster $p'\in\mathcal{P}$ contains exactly one vertex in $p$. Each cluster is a sub path of $p$. Suppose $p'=(v_{a},v_{a+1},..,v_{a+\ell})$, each vertex $v_i\in p'$ also maintain the value $s[v_i]=\sum_{a\le j\le i}x[v_j]$. Initially the value is simply $x[v_i]$. 
			
			We run several phases, at each phase, each cluster toss a coin, getting head or tail. In each phase, tail clusters merge to head clusters if they share an edge $(v_i,v_{i+1})$, where $v_i$ is the head cluster and $v_{i+1}$ is the tail cluster. One can see that each tail cluster can merge to at most $1$ head cluster, and each head cluster can merge at most $1$ tail cluster. $v_i$ sends the value $s[v_i]$ to all the vertices in the cluster containing $v_{i+1}$, and they add this value to there results. Communication inside each cluster happened by broadcasting in each cluster if the cluster has size at most $d$, or broadcasting through the whole network if it has size more than $d$. Since the path has length $k$, the number of clusters that broadcast through the whole network is bounded by $k/d$. Since at each phase, each edge has constant probability to merge two clusters (and become an edge inside a cluster), after $O(\log n)$ phases, there are no edge connecting two different cluster. 
		\end{proof}

		\begin{proof}[Proof of Lemma~\ref{lem:FindPath}]
			The idea is to combine directed rooted trees while maintain the tree properties. We first prove the following frequently used claim: given several vertex disjoint rooted directed trees inside $H$, each vertex $v$ knows the subtree size $s[v]$ inside each tree, then given each vertex $v$ a number $x[v]$, with dilation $\tilde{O}(d+D)$ and congestion $\tOh(N/d)$, each vertex get $\sum_{v'\in T_v}x[v']$, where $T_v$ is the subtree rooted at $v$. 
			
			To prove the claim, we use the idea of heavy-light tree decomposition and Lemma~\ref{lem:pushlabel}. We first decompose each tree into paths: each vertex build a directed edge to one of its child with largest subtree size. The decomposition has the property that every path from a root to a leaf will go through at most $\log N$ paths. To see this, consider each time the path from the root to a leaf leaves a path, since a vertex always point to the child with the largest subtree size, the child without the edge from the parent must have at most half of the subtree size. Thus, we only need to run the algorithm in Lemma~\ref{lem:pushlabel} $O(\log n)$ steps: initially, each vertex get the value $x[v]$, and each path calculate $s[v]$ which is the sum of $x[v']$ for all $v'$ after $v$ on this path. At the beginning of any latter steps, each vertex $v$ send the value $s[v]$ to its parent if $v$ is the beginning of some path, i.e., the parent of $v$ do not have an edge pointing to $v$. Then all vertices $v$ set the receive value as $x[v]$ and redo Lemma~\ref{lem:pushlabel}. For each path, if its length is at most $d$, then we do calculation just by pipeline on this path; otherwise we use the Lemma~\ref{lem:pushlabel} to cause a dilation $\tilde{O}(d)$ and congestion $\tilde{O}(k/d)$, where $k$ is the length of the path. Since $H$ has $N$ vertices, there are at most $N/d$ such instance, while the sum of congestion of all instance is bounded by $\tilde{O}(N/d)$. Thus, the claim is proved.
			
			Now we proceed to prove the lemma. We maintain clusters $\mathcal{S}$, initially each cluster $S\in\mathcal{S}$ contains exactly one vertex in $V'$. Each cluster maintain a rooted directed tree, and each vertex in the tree know the size of the subtree rooted on the vertex. We run several phases, at each phase, each cluster tosses a coin, getting head or tail. In each phase, tail clusters merge to head clusters if they share an edge in $G[V']$. Now we consider how to maintain the tree information. Suppose the head cluster is $T$, and there is a tail cluster $T'$ want to merge to $T$ through the edge $(u,v)$ with $u\in T,v\in T'$. $T'$ first reverse all the edges from $v$ to the root of $T'$, this can be done by applying the claim, putting value $1$ on $v$ and value $0$ on any other vertices. All the vertices $v'$ on the path change its subtree value from $s[v']$ to $s[r]-s[v']+1$, where $s[r]$ is the size of the tree $T'$. One can see that this maintains the correct subtree size inside $T'$. Now we want to maintain the correct subtree size inside $T$. $v$ sends the tree size of $T'$ to $u$. Now each leaf of $T$ that hangs a merged subtree get the value of the tree size. $T$ use the claim to add these value to its original subtree size for each vertex. All the communication happens simultaneous, with the same argument: if a tree $T$ has size at most $d$, then all the properties are calculated through a pipeline inside $T$, otherwise use the claim and cause a dilation $\tOh(d+D)$ and congestion $\tOh(|T|/d)$, while the sum of all $|T|$ is bounded by $N$. 
			
			Since each edge become an edge inside a cluster with constant probability at each phase, after $O(\log n)$ phases, $V'$ becomes a whole cluster since $H[V']$is connected. Each phase cause a dilation $\tOh(d+D)$ and congestion $\tOh(N/d)$, the total dilation and congestion are $\tOh(d+D)$ and $\tOh(N/d)$. Now we get a rooted directed tree $T$ on $V'$. To find the path from $s$ to $t$, we can use the claim to find the path from $s$ to the root, and from the root to $t$. 
		\end{proof}
		
		\begin{proof}[Proof of Lemma~\ref{lem:FindCluster}]
			Each vertex in $V\backslash V'$ broadcast tokens to build a BFS tree: initially only vertices in $V\backslash V'$ are activate, on each round all active vertices send a token to all its neighbors if it haven not done so. At the end of each round, if a vertex that have not been activated receive a token (probability more than one token), then it joins the BFS tree of an arbitrary vertex that sent it the token, then become activate. The procedure has dilation $O(D)$ and congestion $O(1)$, since each active vertices only sends once, and each vertex has distance at most $D$ to a vertex in $V\backslash V'$.
			
			Now we get a partition of $V$, each part is a tree rooted at a vertex $v\in V\backslash V'$ with depth at most $D$, we denote it as $T_v$. Each subtree rooted at a child of $v$ on $T_v$ is a cluster $S$. All $S$ form a partition of $V'$, denoted as $\mathcal{S}$, and $G[S]$ is a rooted tree with depth $D$. Now we want to combine clusters in order to increase the number of vertices in each cluster. Each cluster $S$ also maintain a center set $Centers[S]$, initially contain $v$, where $S$ is the cluster which is a subtree of a child of $v$. We will maintain the property that $H_S=\{(u,v)\mid u\in S,v\in S\cup Centers[S]$ has diameter at most $O(kD)$. 
			
			The algorithm runs for several phases, in each phase, each cluster $S$ with $|S|<x$ tosses a coin, getting tail or head with equal probability $1/2$. If there exists an edge $(u,v)$ where $u,v$ are in different clusters $S_u,S_v$, and $S_u$ have head while $S_v$ have tail, then $S_v$ will merge to $S_u$. They merge there center set $Centers[S_u],Centers[S_v]$ as well as their vertex set. Notice that each vertex in any cluster still have distance at most $D$ to one of its center. Thus, as long as a cluster $S$ is connected, the set $H_S$ has diameter at most $O(kD)$, because all the vertices that has distance to a certain center has mutual distance at most $2D$. Since an edge connecting two small cluster has constant probability to merge at one phase, there are at most $O(\log n)$ phases with high probability. Each phase contains a broadcasting with dilation $O(kD)$ and congestion $O(k)$.
			
			At last, all the edge $(u,v)$ in $H[V']$ has the following property: either $u,v$ are in the same cluster, or one of the cluster containing $u,v$ has size at least $x$. Since merging happens on two clusters with size both at most $x$, if a final cluster have been merged during the algorithm, it must have size $\Theta(x)$; otherwise, it has never been merged, which means it remains a rooted tree with depth $D$. Now suppose $s$ is in cluster $S$. If $|S|=\Theta(x)$ then we are done; otherwise, either $S$ is an initial cluster, or $|S|=o(x)$ and one of the neighbor $S'$ of $S$ has size $\Omega(x)$. In the later case, we will include both $S$ and some part of $S'$ into $V''$. If $|S'|=\Theta(x)$, then we are done, otherwise $S'$ is an initial cluster. So the only thing left is the following problem: given a rooted tree $T$ with depth $D$, number of vertices $\Omega(x)$ and a vertex $s\in T$, find a connected components in $T$ containing $s$ with size $\Theta(x)$. With out loss of generality, we can assume $s$ is the root of $T$, while the depth is still bounded by $O(D)$. We first compute for each vertex $v$ in $T$ the subtree size $s[v]$, by a pipeline on $T$. Then we compute the preorder traversal number for each vertex in $T$ in the following way: suppose a vertex $v$ has the preorder traversal number $p[v]$, and it has $\ell$ children $v_1,...,v_{\ell}$, then $v_i$ get the preorder traversal number $\sum_{1\le j<i}s[v_j]+p[v]+1$. Start from the root with preorder traversal number $1$, the preorder traversal number for layer 2 vertices in $O(1)$ rounds, then layer 3, 4... The total dilation is $O(D)$. 
		\end{proof}
	\end{document}